\documentclass{sig-alternate}
\usepackage{amsmath,amsfonts,times,url}
\usepackage{color}

\newcommand{\final}{1}

\ifnum\final=0
\newcommand{\mnote}[1]{{\tiny \bf {{#1}}}}
\else
\newcommand{\mnote}[1]{}
\fi

\usepackage[small,it]{caption}

\newcommand{\sampnum}{\mathcal{A}}
\newcommand{\AAA}{\mathcal{A}}

\newcommand{\remove}[1]{}
\newcommand{\ignore}[1]{}

\newcommand{\paren}[1]{\left( {#1} \right)}

\newtheorem{theorem}{Theorem}
\newtheorem{lemma}[theorem]{Lemma}

\newtheorem{proposition}[theorem]{Proposition}

\newtheorem{myexample}{Example}

\newtheorem{myremark}{Remark}

\newtheorem{definition}[theorem]{Definition}

\newcommand{\thmref}[1]{Theorem~\ref{thm:#1}}

\newcommand{\defref}[1]{Definition~\ref{def:#1}}

\newcommand{\secref}[1]{Section~\ref{sec:#1}}

\newcommand{\mypar}[1]{\smallskip\noindent {\bf {#1}.}}

\newcommand{\eps}{\epsilon}

\newcommand{\from}{\leftarrow}

\DeclareMathSymbol{\erert}{\mathbin}{AMSb}{"50}

\newcommand{\A}{{\cal A}}

\newcommand{\sd}[1]{\mathbf{SD}\paren{{#1}}}

\newcommand{\ff}{{S}}

\renewcommand{\S}{\ff}

\newcommand{\defeq}{\stackrel{{\mbox{\tiny def}}}{=}}

\newcommand{\beq}{\begin{equation}}
\newcommand{\eeq}{\end{equation}}
\newcommand{\bml}{{\begin{multline}}}
\newcommand{\eml}{{\end{multline}}}
\DeclareMathSymbol{\Q}{\mathbin}{AMSb}{"51}

\ifnum\final=1
\newcommand{\mynote}[1]{\marginpar{\tiny\sf #1}}
\else
\newcommand{\mynote}[1]{}
\fi

\newcommand{\D}{{\mathcal{D}}}

\newenvironment{CompactItemize}{\begin{itemize}}{\end{itemize}}
\newenvironment{CompactEnumerate}{
  \begin{list}{\arabic{enumi}.}{%
      \usecounter{enumi}
      \setlength{\leftmargin}{12pt}
      \setlength{\itemindent}{3pt}
      \setlength{\topsep}{3pt}
      \setlength{\itemsep}{1pt}
      }}
  {\end{list}}

\begin{document}
%
% --- Author Metadata here ---
%\conferenceinfo{SIGMOD'08}{'08 Vancouver, BC, Canada}
%\CopyrightYear{2008} % Allows default copyright year (2002) to be over-ridden - IF NEED BE.
%\crdata{978-1-59593-753-7/08/0003}  % Allows default copyright data (X-XXXXX-XX-X/XX/XX) to be over-ridden.
% --- End of Author Metadata ---

\title{Composition Attacks and Auxiliary Information in Data Privacy}
%\subtitle{Paper ID: EIS-113}
%\titlenote{A full version of this paper is available as
%\textit{Author's Guide to Preparing ACM SIG Proceedings Using
%\LaTeX$2_\epsilon$\ and BibTeX} at
%\texttt{www.acm.org/eaddress.htm}}}
%
% You need the command \numberofauthors to handle the "boxing"
% and alignment of the authors under the title, and to add
% a section for authors number 4 through n.
%
% Up to the first three authors are aligned under the title;
% use the \alignauthor commands below to handle those names
% and affiliations. Add names, affiliations, addresses for
% additional authors as the argument to \additionalauthors;
% these will be set for you without further effort on your
% part as the last section in the body of your article BEFORE
% References or any Appendices.

\numberofauthors{3}

% You can go ahead and credit authors number 4+ here;
% their names will appear in a section called
% "Additional Authors" just before the Appendices
% (if there are any) or Bibliography (if there
% aren't)
%
% Put no more than the first THREE authors in the \author command
\author{
% You can go ahead and credit any number of authors here,
% e.g. one 'row of three' or two rows (consisting of one row of three
% and a second row of one, two or three).
%
% The command \alignauthor (no curly braces needed) should
% precede each author name, affiliation/snail-mail address and
% e-mail address. Additionally, tag each line of
% affiliation/address with \affaddr, and tag the
% e-mail address with \email.
%
% 1st. author
\alignauthor
Srivatsava Ranjit Ganta
%%\titlenote{Dr.~Trovato insisted his name be first.}
			\\
       \affaddr{Penn State University}\\
       \affaddr{University Park, PA 16802}\\
       \email{ranjit@cse.psu.edu}
\alignauthor
Shiva Kasiviswanathan
%%\titlenote{The secretary disavows
%%any knowledge of this author's actions.}
			\\
       \affaddr{Penn State University}\\
       \affaddr{University Park, PA 16802}\\
%%       \affaddr{San Jose, CA 95123}\\
       \email{kasivisw@cse.psu.edu}
\alignauthor
Adam Smith
%%\titlenote{The secretary disavows
%%any knowledge of this author's actions.}
			\\
       \affaddr{Penn State University}\\
       \affaddr{University Park, PA 16802}\\
%%       \affaddr{San Jose, CA 95123}\\
       \email{asmith@cse.psu.edu}   
%\alignauthor
%Raj Acharya
%%%\titlenote{The secretary disavows
%%%any knowledge of this author's actions.}
%			\\
%       \affaddr{Penn State University}\\
%       \affaddr{University Park, PA 16802}\\
%%%       \affaddr{San Jose, CA 95123}\\
%       \email{acharya@cse.psu.edu}        
}
\maketitle
\begin{abstract}
Privacy is an increasingly important aspect of data publishing. Reasoning about privacy, however, is fraught with pitfalls. One of the most significant is the auxiliary information (also called external knowledge, background knowledge, or side
information) that an adversary gleans from other channels such as the web, public records, or domain knowledge. This paper explores how one can reason about privacy in the face of rich, realistic sources of auxiliary information. Specifically, we investigate the effectiveness of current anonymization schemes in preserving privacy when multiple organizations {\em independently} release anonymized data about overlapping populations.

\begin{CompactEnumerate}
\item We investigate {\em composition attacks}, in which an adversary uses independent anonymized releases to breach privacy. We explain why recently proposed models of limited auxiliary information fail to capture composition attacks. Our experiments demonstrate that even a simple instance of a composition attack can breach privacy in practice, for a large class of currently proposed techniques. The class includes $k$-anonymity and several recent variants.

\item On a more positive note, certain randomization-based notions of privacy (such as differential privacy) provably resist composition attacks and, in fact, the use of arbitrary side information. This resistance enables ``stand-alone" design of anonymization schemes, without the need for explicitly keeping track of other releases. 
We provide a precise formulation of this property, and prove that an important class of  relaxations of differential privacy also satisfy the property. This significantly enlarges the class of protocols known to enable modular design.
\end{CompactEnumerate}
\end{abstract}

\section{Introduction}
Privacy is an increasingly important aspect of data publishing. 
%Large volumes of personal and sensitive data are collected and archived by health networks, government agencies, search engines, social networking websites, and other organizations.
%
The potential social benefits of analyzing large collections of personal information (census data, medical records, social networks) are significant. At the same time, the release of information from such repositories %with sensitive data 
can be devastating to the privacy of individuals or organizations~\cite{nyt06aol}. 
The challenge is therefore to discover and release the global characteristics of these databases without compromising the privacy of the individuals whose data they contain. 

Reasoning about privacy, however, is fraught with pitfalls. %The problem has been studied in statistics since the 1970's and more recently in the database, data mining, and cryptography literature (under various names: {\em statistical disclosure control, privacy-preserving data mining} and {\em private data analysis}, to name a few).
 One of the most significant difficulties is the auxiliary information (also called external knowledge, background knowledge, or side
information) that an adversary gleans from other channels such as the web or public records. For example, simply removing obviously identifying information such as names and address does not suffice to protect privacy since the remaining information (such as zip code, gender and date of birth~\cite{Sweeney}) may still identify a person uniquely when combined with auxiliary information (such as voter registration records). 
%In such an attack, the anonymized record corresponding to an individual is identified via seemingly innocuous attributes (such as zip code, gender and date of birth~\cite{Sweeney}) using externally available information (such as voter registration records). 
Schemes that resist such linkage have been the focus of extensive investigation, starting with work on publishing contingency tables~\cite{jos-si}, and more recently, in a line of techniques based on ``$k$-anonymity'' 
\cite{Sweeney}.

This paper explores how one can reason about privacy in the face of rich, realistic sources of auxiliary information. This follows lines of work in both the data mining \cite{MGKV06,MartinKMGH07,ChenRL07} and cryptography~\cite{DiNi03,Dwork06} communities that have sought principled ways to incorporate {\em unknown} auxiliary information into anonymization schemes. 
Specifically, we investigate the effectiveness of current anonymization schemes in preserving privacy when multiple organizations {\em independently} release anonymized data about overlapping populations. We show new attacks on some schemes and also deepen the current understanding of schemes known to resist such attacks. Our results and their relation to previous work are discussed below.

Schemes that retain privacy guarantees in the presence of independent releases are said to {\em compose securely}. The terminology, borrowed from cryptography (which borrowed, in turn, from software engineering), stems from the fact that schemes which compose securely can be designed in a stand-alone fashion without explicitly taking other releases into account. %and yet safely used as part of a larger data management system. 
Thus, understanding independent releases is essential for enabling modular design. In fact, one would like schemes that compose securely not only with independent instances of themselves, but with {\em arbitrary external knowledge}. We discuss both types of compositions in this paper.

%<<<<<<< .mine
%The dual problem to designing schemes with good composition properties is the design of attacks that exploit such information. We call these {\em composition attacks}, again following cryptographic terminology. A simple example of such an attack, in which two hospitals with overlapping patient populations publish anonymized medical data, is presented below. 
%Composition attacks highlight a realistic and important class of vulnerabilities. 
%As privacy preserving data publishing becomes more commonly deployed, it is increasingly difficult to keep track of all the organizations that publish anonymized summaries involving a given individual or entity and schemes that are vulnerable to composition attacks will become increasingly difficult to use safely.%\an{say more here: really two points, 1. much more data to protect, and 2. much more side info for attacker.}
%=======
The dual problem to designing schemes with good composition properties is the design of attacks that exploit such information. We call these {\em composition attacks}.%, again following cryptographic terminology. 
A simple example of such an attack, in which two hospitals with overlapping patient populations publish anonymized medical data, is presented below. 
Composition attacks highlight a realistic and important class of vulnerabilities. 
As privacy preserving data publishing becomes more commonly deployed, it is increasingly difficult to keep track of all the organizations that publish anonymized summaries involving a given individual or entity and schemes that are vulnerable to composition attacks will become increasingly difficult to use safely.%\an{say more here: really two points, 1. much more data to protect, and 2. much more side info for attacker.}
%>>>>>>> .r437

\subsection{Contributions}
Our contributions are summarized briefly in the abstract, above, and discussed in more detail in the following subsections.
\subsubsection{\hspace*{-6pt}Composition Attacks on Partition-based Schemes}

We introduce composition attacks and study their effect on a popular class of partitioning-based anonymization schemes. Very roughly, computer scientists have worked on two broad classes of anonymization techniques. {\em Randomization-based} schemes introduce uncertainty either by randomly perturbing the raw data (a technique called {\em input perturbation}, {\em randomized response}, e.g., ~\cite{W65,agrawal-ppdm, evfimievski02}), or {\em post-randomization}, e.g.,~\cite{HH02}),  or by injecting randomness into the algorithm used to analyze the data (e.g.,~\cite{BDMN05,MT07}). %The intuition is that the added randomness makes it difficult for an adversary viewing the public results to infer concrete information about an individual. 
{\em Partition-based schemes} cluster the individuals in the database into disjoint groups satisfying certain criteria (for example, in $k$-anony\-mity \cite{Sweeney}, each group must have size at least $k$). 
For each group, certain exact statistics are calculated and published. 
Partition-based schemes include $k$-anonymity \cite{Sweeney} as well as several recent variants, e.g., \cite{MGKV06,LiLV07,XiaoT07,MartinKMGH07,ChenRL07}.
%$K$-anonymity~\cite{SamaratiandSweeney1998}, $\ell$-diversity~\cite{MGKV06} and other work in this line~\cite{LiLV07} achieve partitioning through generalization and aggregation techniques. On the other hand, techniques such as~\cite{AggarwalFKKPTZ06},~\cite{Ferrer2002} achieve this by clustering the data. %Partitioning based solutions are mainly applied to non-interactive scenarios where the data needs to be published after anonymization.

Because they release exact information, partition-based schemes seem especially vulnerable to composition attacks. In the first part of this paper we study a simple instance of a composition attack called an {\em intersection attack}. We observe that the specific properties of current anonymization schemes make this attack possible, and we evaluate its success empirically.

\mypar{Example} Suppose two hospitals $H_1$ and $H_2$ in the same city release anonymized patient-discharge information. Because they are in the same city, some patients may visit both hospitals with similar ailments. Tables~\ref{tab:example}(a) and~\ref{tab:example}(b) represent (hypothetical) independent $k$-anonymizations of the discharge data from $H_1$ and $H_2$ using $k=4$ and $k=6$, respectively. The sensitive attribute here is the patient's medical condition. It is left untouched.  The other attributes, deemed non-sensitive, are generalized  (that is, replaced with aggregate values), so that within each group of rows, the vectors if non-sensitive attributes are identical. 
If Alice's employer knows that she is 28 years old, lives in zip code 13012 and recently visited both hospitals, then he can attempt to locate her in both anonymized tables. Alice matches four potential records in $H_1$'s data, and six potential records in $H_2$'s. However, the only disease that appears in both matching lists is AIDS, and so Alice's employer learns the reason for her visit.

\begin{table}[htb]{\scriptsize\sf
\centering
		\begin{tabular}{|c||c|c|c||c||} \hline
		  & \multicolumn{3}{c||}{\textbf{Non-Sensitive}} & \textbf{Sensitive} \\ \hline
			&\textbf {Zip code}	&\textbf {Age}	&\textbf {Nationality} &\textbf {Condition} \\ \hline
			1	&130** &$<$30	&* &AIDS\\ 
			2 &130** &$<$30  &* &Heart Disease\\ 
			3	&130** &$<$30	&*	&Viral Infection\\ 
			4 &130** &$<$30	&*	&Viral Infection\\ \hline
			5	&130** &$\geq$40	&*	&Cancer\\ 
			6	&130** &$\geq$40	&*	&Heart Disease\\ 
			7	&130** &$\geq$40	&*	&Viral Infection\\ 
			8	&130** &$\geq$40	&*	&Viral Infection\\ \hline
			9	&130** &3*	&*	&Cancer\\ 
			10 &130** &3*	&*	&Cancer\\ 
			11 &130** &3*	&*	&Cancer\\ 
			12 &130** &3*	&*	&Cancer\\ \hline
		\end{tabular}\\
		(a)

		\begin{tabular}{|c||c|c|c||c||} \hline
		  & \multicolumn{3}{c||}{\textbf{Non-Sensitive}} & \textbf{Sensitive} \\ \hline
			&\textbf {Zip code}	&\textbf {Age}	&\textbf {Nationality} &\textbf {Condition} \\ \hline
			1	&130** &$<$35	&* &AIDS\\ 
			2 &130** &$<$35  &* &Tuberculosis\\ 
			3	&130** &$<$35	&*	&Flu\\ 
			4 &130** &$<$35	&*	&Tuberculosis\\
			5	&130** &$<$35	&*	&Cancer\\ 
			6 &130** &$<$35	&*	&Cancer\\ \hline
			7 &130** &$\geq$35	&*	&Cancer\\ 
			8 &130** &$\geq$35	&*  &Cancer\\ 
			9	&130** &$\geq$35	&*	&Cancer\\ 
			10 &130** &$\geq$35	&*	&Tuberculosis\\ 
			11 &130** &$\geq$35	&*	&Viral Infection\\ 
			12 &130** &$\geq$35	&*	&Viral Infection\\ \hline
		\end{tabular}
\\
(b)

} 
%\end{array}$
\caption{%(a) Patient Data from H1 (b) Patient Data from H2 
%(c) 4-anonymous $H_1$ Patient Data (d) 6-anonymous $H_2$ Patient Data} 
\small\rm
A simple example of a composition attack. Tables (a) and (b) are 4-anonymous (respectively, 6-anonymous) patient data from two hypothetical hospitals. If an Alice's employer knows that she is 28, lives in zip code 13012  and visits both hospitals, he learns that she has AIDS. }
\label{tab:example}
\end{table} 
 \mypar{Intersection Attacks}
The above example relies on two properties of the partition-based anonymization schemes: {\bf (i)  Exact sensitive value disclosure:} the ``sensitive'' value corresponding to each member of the group is published exactly;  and {\bf (ii) Locatability:} given any individual's non-sensitive values (non-sensitive values are exactly those that are assumed to be obtainable from other, public information sources) one can locate the group in which individual has been put in. 
Based on these properties, an adversary can narrow down the set of possible sensitive values for an individual by intersecting the sets of sensitive values present in his/her groups from multiple anonymized releases.  

Properties (i) and (ii) turn out to be widespread. The exact disclosure of sensitive value lists is a design feature common to all the schemes based on $k$-anonymity: preserving the exact distribution of sensitive values is important, and so no recoding is usually applied. 
Locatability is less universal, since it depends on the exact choice of clustering algorithm (used to form groups) and the recoding applied to the non-sensitive attributes. However, some schemes always satisfy locatability by virtue of their structure (e.g., schemes that
recursively partition the data set along the lines of a hierarchy that is subsequently used for generalization~\cite{LeFevreDR06,LeFevreDR06workload}). For other schemes, locatability is not perfect but our experiments suggest that using simple heuristics one can locate a individual's group with high probability.

\remove{Even with these properties, it is difficult to come up with a theoretical model that predicts the effectiveness of intersection attacks. For example, if the sensitive values of the members of a group could be assumed to be statistically independent of their non-sensitive attribute values, then a simple birthday-paradox-style analysis would yield reasonable bounds. However, the partitioning techniques, which consider both types of attributes, creates dependencies that are hard to model analytically. }

Even with these properties, it is difficult to come up with a theoretical model for intersection attacks because the partitioning techniques generally create dependencies that are hard to model analytically. However, if the sensitive values of the members of a group could be assumed to be statistically independent of their non-sensitive attribute values, then a simple birthday-paradox-style analysis would yield reasonable bounds.

\mypar{Experimental Results} Instead, we evaluated the success of intersection attacks empirically. We ran the intersection attack on two popular census databases anonymized using partition-based schemes. We evaluated the severity of such an attack by measuring the number of individuals who had their sensitive value revealed. Our experimental results confirm that partitioning-based anonymization schemes including $k$-anonymity and its recent variants, $\ell$-diversity and $t$-closeness, are indeed vulnerable to intersection attacks.   \secref{experiments} elaborates our methodology and results. 

\mypar{Related Work on Modeling Background Knowledge}
It is important to point out that the partition-based schemes in the literature were not designed to be used in contexts where independent releases are available. Thus, we do not view our results as pointing out a flaw in these schemes, but rather as directing the community's attention to an important direction for future work. 

It is equally important to highlight the progress that has already been made on modeling sophisticated background knowledge in partition-based schemes.  One line has focused on taking into account other, {\em known} releases, such as previous publications by the same organization (``sequential'' releases, \cite{WangF06,ByunSBL06,XiaoT07}) and multiple views of the same data set \cite{YaoWJ05}. Another line has considered incorporating knowledge of the clustering algorithm used to group individuals \cite{WongFWP07}. Most relevant to this paper are works that have sought to model {\em unknown} background knowledge. Martin {\em et al.}~\cite{MartinKMGH07} and Chen {\em et al.}~\cite{ChenRL07} provide complexity measures for an adversary's side information (roughly, they measure the size of the smallest formula within a CNF-like class that can encode the side information). Both works design schemes that provably resist attacks based on side information whose complexity is below a given threshold.  

Independent releases (and hence composition attacks) fall outside the models proposed by these works. The sequential release models do not fit because they deal assume the other releases are known to the anonymization algorithm. The complexity-based measures do not fit because independent releases appear to have complexity that is linear in the size of the data set.

\subsubsection{\hspace*{-6pt} Composing Randomization-based Schemes}
\label{sec:bayes-intro}

Composition attacks appear to be difficult to reason about, and it is not initially clear whether it is possible at all to design schemes that resist such attacks. Even {\em defining} composition properties precisely is tricky in the presence of malicious behavior (for example, %definitions of composability for cryptographic protocols are a topic of extensive research--
see \cite{Lindell-book} for a recent survey about composability of cryptographic protocols).  Nevertheless, a significant family of anonymization definitions do provide   guarantees against composition attacks, namely schemes that satisfy  {\em differential privacy}~\cite{DMNS06}. 
%and several relaxations~\cite{DKMMN06,CM06,NRS07,MKAGV08}. (For clarity, we will refer to the simplest definition \cite{DMNS06} as {\em strict} differential privacy.)
Recent work has greatly expanded the applicability of  differential privacy and its relaxations, both in the theoretical~\cite{DwNi04,BDMN05,DMNS06,BCDKMT07,MT07} and applied \cite{EGS03,AH05,MKAGV08} literature. However, certain recently developed techniques such as sampling~\cite{CM06}, instance-based noise addition~\cite{NRS07} and data synthesis~\cite{MKAGV08} appear to require relaxations of the definition.

It is simple to prove that both the strict and relaxed variants of differential privacy  compose well (see~\cite{DKMMN06,NRS07,MT07}). Less trivially, however, one can prove that strictly differentially-private algorithms also provide meaningful privacy in the presence of {\em arbitrary side information} (Dwork and McSherry, \cite{Dwork06}). In particular, these schemes compose well even with completely different anonymization schemes. 

It is natural to ask if there are weaker definitions which provide similar guarantees. Certainly not all of them do: one natural relaxation of differential privacy, which replaces the multiplicative distance used in differential privacy with total variation distance, fails completely to protect privacy (see example 2 in \cite{DMNS06}).%\an{Mention other types of randomization that fail?}

In this paper, we prove that two important relaxations of differential privacy do, indeed, resist arbitrary side information. First, we provide a Bayesian formulation of differential privacy which makes its resistance to arbitrary side information explicit. Second, we prove that the relaxed definitions of \cite{DKMMN06,MKAGV08} still imply the Bayesian formulation.  The proof is non-trivial, and relies on the ``continuity'' of Bayes' rule with respect to certain distance measures on probability distributions. Our result means that the recent techniques mentioned above~\cite{DKMMN06,CM06,NRS07,MKAGV08} can be used modularly with the same sort of assurances as in the case of strictly differentially-private algorithms. 

\section{Partition-Based Schemes} \label{sec:definitions}

%We now provide basic definitions of the partitioning based techniques. 
Let $D$ be a multiset of tuples where each tuple corresponds to an individual in the database. Let $R$ be an anonymized version of $D$. 
%Let $T = {t_1, t_2, \ldots, t_p}$ be a sensitive database table with attributes $A_1, \ldots, A_m$. 
%Let $R = {t_1, t_2, \ldots, t_p}$ be an anonymized release of a sensitive database. It is a multiset of tuples $t_1, t_2, \ldots, t_p$ where each tuple $t$ corresponds to an individual in the database. 
From this point on, we use the terms tuple and individual interchangeably, unless the context leads to ambiguity. %Let $D$ be the original database corresponding to the anonymized release $R$. 
Let $A = A_1,A_2, \ldots, A_r$ be a collection of attributes and $t$ be a tuple in $R$; we use the notation $t[A]$ to denote $(t[A_1], \ldots, t[A_r])$ where each $t[A_i]$ denotes the value of attribute $A_i$ in table $R$ for $t$.  

%which is a projection of tuple $t$ onto attributes in $A$.  
%Let $t[A_i]$ denote the value of attribute $A_i$ in table $R$ for a tuple $t$. 

In partitioning-based anonymization approaches, there exists a division of data attributes into two classes, \emph{sensitive attributes} and \emph{non-sensitive attributes}. A sensitive attribute is one whose value and an individual's association with that value should not be disclosed. %In the example given in the previous section, the medical condition of the patient is a sensitive attribute. 
%Previous work in this line typically assumed that there is a single sensitive attribute. 
All attributes other than  the sensitive attributes are non-sensitive attributes.

\begin{definition}[Quasi-identifier] A set of non-sens\-itive attributes $\{Q_1,\ldots,Q_{r}\}$ is called a  \textbf{quasi-identifier} if there is at least one individual in the original sensitive database $D$ who can be uniquely identified by linking these attributes with auxiliary data. \end{definition}

\remove{In the example given in the previous section, the set of attributes $\{$zip code, sex, nationality$\}$ form the quasi-identifier.}
% adam 2/29/08: raw data was not given so this makes no sense
Previous work in this line typically assumed that all the attributes in the database other than the sensitive attribute form the quasi-identifier.

\begin{definition}[Equivalence Class] An \textbf{equivalence cla\-ss}  for a table $R$ with respect to attributes in $A$ is the set of all tuples $t_1, t_2, \ldots, t_i \in R$ for which the projection of each tuple onto attributes in $A$ is the same, i.e., $t_1[A] = t_2[A] \ldots = t_i[A]$. \end{definition}

Partition-based schemes cluster individuals into groups, and then recode (\emph{i.e.,} generalize or change) the non-sensitive values so that each group forms an equivalence class with respect to the quasi-identifiers. Sensitive values are not recoded. Different criteria are used to decide how, exactly, the groups should be structured. The most common rule is  $k$-anonymity, which requires that each equivalence class contain at least $k$ individuals.
\remove{We first define $k$-anonymity which is the basic definition of privacy used in partitioning-based anonymization schemes. Informally, $k$-anonymity divides a sensitive database into a set of partitions based on the non-sensitive attributes, such that, the \emph{minimum partition-size} is at least $k$.}

\begin{definition}[$k$-anonymity] A release $R$ is  $k$\textbf{-an\-onym\-ous} if for every tuple $t \in R$, there exist at least $k-1$ other tuples $t_1, t_2, \ldots, t_{k-1} \in R$ such that $t[A] = t_1[A] = \ldots = t_{k-1}[A]$ for every collection $A$ of attributes in quasi-identifier. \end{definition}
In our experiments we also consider two extensions to $k$-anonymity.
%We now briefly define the extensions proposed to the original $k$-anonymity definition that we consider in this study. 

\begin{definition}[Entropy $\ell$-diversity] For an equivalence class $E$, let $S$ denote the domain of the sensitive attributes, and $p(E,s)$ is the fraction of records in $E$ that have sensitive value $s$, then $E$ is \textbf{$\ell$-diverse} if: $$-\sum_{s \in S}{p(E,s)\log(p(E,s))} \geq \log l  \,.$$ A table is $\ell$-diverse if all its equivalence classes are $\ell$-diverse.\end{definition}

\begin{definition} [$t$-closeness] An equivalence class $E$ is \textbf{$t$-close} if the distance between the distribution of a sensitive attribute in this class and distribution of the attribute in the whole table is no more than a threshold $t$. A table is  $t$-close if all its equivalence classes are $t$-close. \end{definition}

\vspace{-4pt}
\mypar{Locatability} 
As mentioned in the introduction, many anonymization algorithms satisfy {\em locatability}, that is, they output tables in which one can locate an individual's  group  based only on his or her {\em non-sensitive} values. 
\remove{
For an individual, given the set of non-sensitive attribute values corresponding to his/her record and a $k$-anonymized release that contains the individual's record, the property that allows an adversary to identify the \emph{equivalence class} into which the individual falls is referred to as locatability.
}

\begin{definition}[locatability]  Let $Q$ be the set of quasi-identifier values of an individual in the original database $D$. Given the $k$-anonymized release $R$ of $D$, the locatability property allows an adversary to identify the set of tuples $\{t_1, \ldots, t_K\}$ in $R$ (where $K \geq k$) that correspond to $Q$.  \end{definition} 

Locatability does not necessarily hold for all partition-based sche\-mes, since it depends on 
the exact choice of clustering algorithm (used to form groups) and the recoding applied to the non-sensitive attributes. However it is widespread. Some schemes {\em always} satisfy locatability by virtue of their structure (\emph{e.g.}, schemes that
recursively partition the data set along the lines of a hierarchy always provide locatability if the attributes are then generalized using the same hierarchy, or if (min,max) summaries are used~\cite{LeFevreDR06,LeFevreDR06workload}). For other schemes, locatability is not perfect but our experiments suggest that using simple heuristics can locate a person's group with good probability. For example, microaggregation~\cite{Ferrer2002} clusters individuals based on Euclidean distance. The vectors of non-sensitive values in each group are replaced by the centroid (i.e., average) of the vectors. The simplest heuristic for locating an individual's group is to choose the group with the closest centroid vector. In experiments on census data, this correctly located approximately 70\% of individuals. 
In our attacks, we always assume locatability. This assumption was also made in previous studies~\cite{Sweeney,MartinKMGH07}.

\subsection{Intersection Attack} \label{sec:intattack}

Armed with these basic definitions, we now proceed to formalize the intersection attack (Algorithm 1). 
\begin{algorithm}
\caption{Intersection attack}
\label{algo1}
\begin{algorithmic}[1]
\STATE $R_1,\ldots,R_n \leftarrow$ $n$ independent anonymized releases 
\STATE $P \leftarrow$ set of overlapping population 

\FOR{each individual $i$ in $P$}
\FOR{$j=1$ to $n$}
\STATE $e_{ij} \leftarrow$  Get\_equivalence\_class$(R_j, i)$\;
\STATE $s_{ij} \leftarrow$  Sensitive\_value\_set$(e_{ij})$\;
\ENDFOR
\STATE $S_i \leftarrow s_{i1} \cap s_{i2} \cap \ldots \cap s_{in} $\;
\ENDFOR
RETURN $S_1,\ldots,S_{|P|}$
\end{algorithmic}
\end{algorithm}

%\subsection{Discussion}

Let $R_1,\ldots,R_n$ be $n$ independent anonymized releases with minimum partition-sizes of $k_1,\ldots,k_n$, respectively. Let $P$ be the overlapping population occurring in all the releases. The function Get\_equivalence\_class returns the equivalence class into which an individual falls in a given anonymized release. The function Sensitive\_value\_set returns the set of (distinct) sensitive values for the members in a given equivalence class.

\begin{definition} [Anonymity] For each individual $i$ in $P$, the anonymity factor pro\-mised by each release $R_j$ is equal to the corresponding minimum partition-size $k_j$.  \end{definition}

However, as pointed out in~\cite{MGKV06}, the actual anony\-mity offered is less than this ideal value and is equal to number of distinct values in each equivalence class. We call this as the \emph{effective anonymity}

\begin{definition}[Effective Anonymity]
For an individual $i$ in $P$, the effective ano\-nymity offered by a release $R_j$ is equal to the number of distinct sensitive values of the partition into which the individual falls into. Let $e_{ij}$ be the equivalence class or partition into which $i$ falls into with respect to the release $R_j$, and let $s_{ij}$ denote the sensitive value set for $e_{ij}$. The effective anonymity for $i$ with respect to the release $R_j$ is: $\mathrm{\it EA}_{ij} = |s_{ij}|\,.$ \end{definition}

For each target individual $i$, $\mathrm{\it EA}_{ij}$ is the {\em effective prior anonymity} with respect to $R_j$ (anonymity before the intersection attack). In the intersection attack, the list of possible sensitive values associated to the target is equal to intersection of all sensitive value sets $s_{ij}$, $j=1,\ldots,n$. So the \emph{effective posterior anonymity} ($\widehat{\mathrm{\it EA}}_i$) for $i$ is: $${\widehat{\mathrm{\it EA}}_i} = |\{\cap s_{ij}\}|, j=1,\ldots,n\,.$$

%Note that for all Individuals $i$ in $P$, definition of set intersection results in, 
%\[ Eff\_A^i_{i}_{\mathrm{\it posterior}} \leq \]

The difference between the effective prior anonymity and effective posterior anonymity quantifies the drop in effective anonymity.
$$\mathrm{\it Anon\_Drop}_{i} = \min_{j=1,\ldots,n}\{{\mathrm{\it EA}}_{ij}\} - \widehat{\mathrm{\it EA}}_i\,.$$

The \emph{vulnerable population} ($\mathrm{\it VP}$) is the number of individuals (among the overlapping population) for whom the intersection attack leads to a positive drop in the effective anonymity.
$$\mathrm{\it VP} = \left\{i \in P \,:\, \mathrm{\it Anon\_Drop}_{i} > 0 \right\}\,.$$
%\begin{definition}[Vulnerable Population]
%The \textbf{vulnerable population} 
%\end{definition}

After performing the sensitive value set intersection, the adversary knows only a possible set of values that each individual's sensitive attribute can take. So, the adversary deduces that with equal probability (under the assumption that the adversary does not have any further auxiliary information) the individual's actual sensitive value is one of the values in the set $\left\{\cap s_{ij}\right\}, j=1,\ldots,n$.  So, the adversaries \emph{confidence level} for an individual $i$ can be defined as:

\begin{definition}[Confidence level $C_i$] For each individual $i$, the confidence level $C_i$ of the adversary in identifying the individual's true sensitive value through the intersection attack is defined as $C_i = \frac{1}{{\widehat{\mathrm{\it EA}}_i}}\,.$
\end{definition}

Now, given some confidence level $C$, we denote by $\mathrm{\it VP}_{C}$ and $\mathrm{\it PVP}_{C}$ the set and the percentage of overlapping individuals for whom the adversary can deduce the sensitive attribute value with a confidence level of at least $C$.
\begin{eqnarray*} & \mathrm{\it VP}_{C} = \left\{i \in P \,:\, C_i \geq C \right\},& \\
&\mathrm{\it PVP}_{C} = \frac{|\mathrm{\it VP}_{C}| \cdot 100 }{|P|}  \,.& \end{eqnarray*}
%\end{definition}

\section{Experimental Results}
\label{sec:experiments}

In this section we describe our experimental study\footnote{The code, parameter settings, and complete results are made available at: \url{http://www.cse.psu.edu/~ranjit/kdd08}.}. %of intersection attacks on partitioning based anonymization schemes. 
The primary goal is to quantify the severity of such an attack on existing schemes. % and an insight into information gained by the adversary. 
Although the earlier works address problems with $k$-anonymization and adversarial background knowledge, to the best of our knowledge, none of these studies deal with attacks resulting from auxiliary independent releases. Furthermore, none of the studies so far have quantified the severity of such an attack. 

\subsection{Setup}
We use three different partitioning-based anonymization techniques to demonstrate the intersection attack: $k$-anonymity, $\ell$-diver\-sity, and $t$-closeness. For $k$-anonymity, we use the Mondrian multidimensional approach proposed in~\cite{LeFevreDR06} and the microaggregation technique proposed in~\cite{Ferrer2002}. For $\ell$-diver\-sity and $t$-closeness, we use the definitions of entropy $\ell$-diversity and $t$-closeness proposed in~\cite{MGKV06} and~\cite{LiLV07}, respectively.

We use two census-based databases from the UCI Machine Learning repository \cite{uci}. %as the experimental data. 
The first one is the Adult database that has been used extensively in the $k$-anonymity based studies. The database was prepared in a similar manner to previous studies \cite{LeFevreDR06,MGKV06} (also explained in Table~\ref{adultdesc}). The resulting database contained individual records corresponding to $30162$ people. The second database is the IPUMS database that contains individual information from the 1997 census studies. We only use a subset of the attributes that are similar to the attributes present in the Adult database to maintain uniformity and to maintain quasi-identifiers. %that are known to identify majority of population (see, e.g., \cite{SamaratiandSweeney1998}). 
The IPUMS database contains individual records corresponding to a total of $70187$ people. This data set was prepared as explained in Table~\ref{ipumsdesc}. 

From both Adult and IPUMS databases, we generate two overlapping subsets (Subset 1 and Subset 2) by randomly sampling individuals without replacement from the total population. We fixed the overlap size to $P = 5000$. For each of the databases, the two subsets are anonymized independently and the intersection attack is run on the anonymization results. All the experiments were run on a Pentium 4 system running Windows XP with 1GB RAM. 

\begin{table}\small
	\centering
		\begin{tabular}{||c|c|c||} \hline
			\textbf{Attribute}	& \textbf{Domain Size}	& \textbf{Class} \\ \hline
			Age	&74	   &Quasi ID \\ 
			Work Class	&7	&Quasi ID \\ 
			Education	&16	&Quasi ID \\ 
			Marital Status	&7	&Quasi ID \\ 
			Race	&5	&Quasi ID \\ 
			Gender	&2   &Quasi ID \\ 
			Native Country	&41	&Quasi ID \\ 
			Occupation	&14 &Sensitive \\ \hline
		\end{tabular}	
		\caption{Description of the Adult census database.}
	\label{adultdesc}
\end{table}

\begin{table}\small
	\centering
		\begin{tabular}{||c|c|c||} \hline
			\textbf{Attribute}	& \textbf{Domain Size}	& \textbf{Class} \\ \hline
			Age	&100	   &Quasi ID \\ 
			Work Class	&5	&Quasi ID \\ 
			Education	&10	&Quasi ID \\ 
			Marital Status	&6	&Quasi ID \\ 
			Race	&7	&Quasi ID \\ 
			Sex	&2   &Quasi ID \\ 
			Birth Place	&113	&Quasi ID \\ 
			Occupation	&247 &Sensitive \\ \hline
		\end{tabular}	
		\caption{Description of the IPUMS census database.}
	\label{ipumsdesc}
\end{table}

\subsection{Severity of the Attack}
Our first goal is to quantify the extent of damage possible through the intersection attack. For this, we consider two possible situations: (i) Perfect breach and (ii) Partial breach.
\subsubsection{\hspace*{-6pt} Perfect Breach}
\begin{figure*}[!htb]
\begin{footnotesize}
\centering $\begin{array}{cc}
\includegraphics[width=225pt, height=150pt]{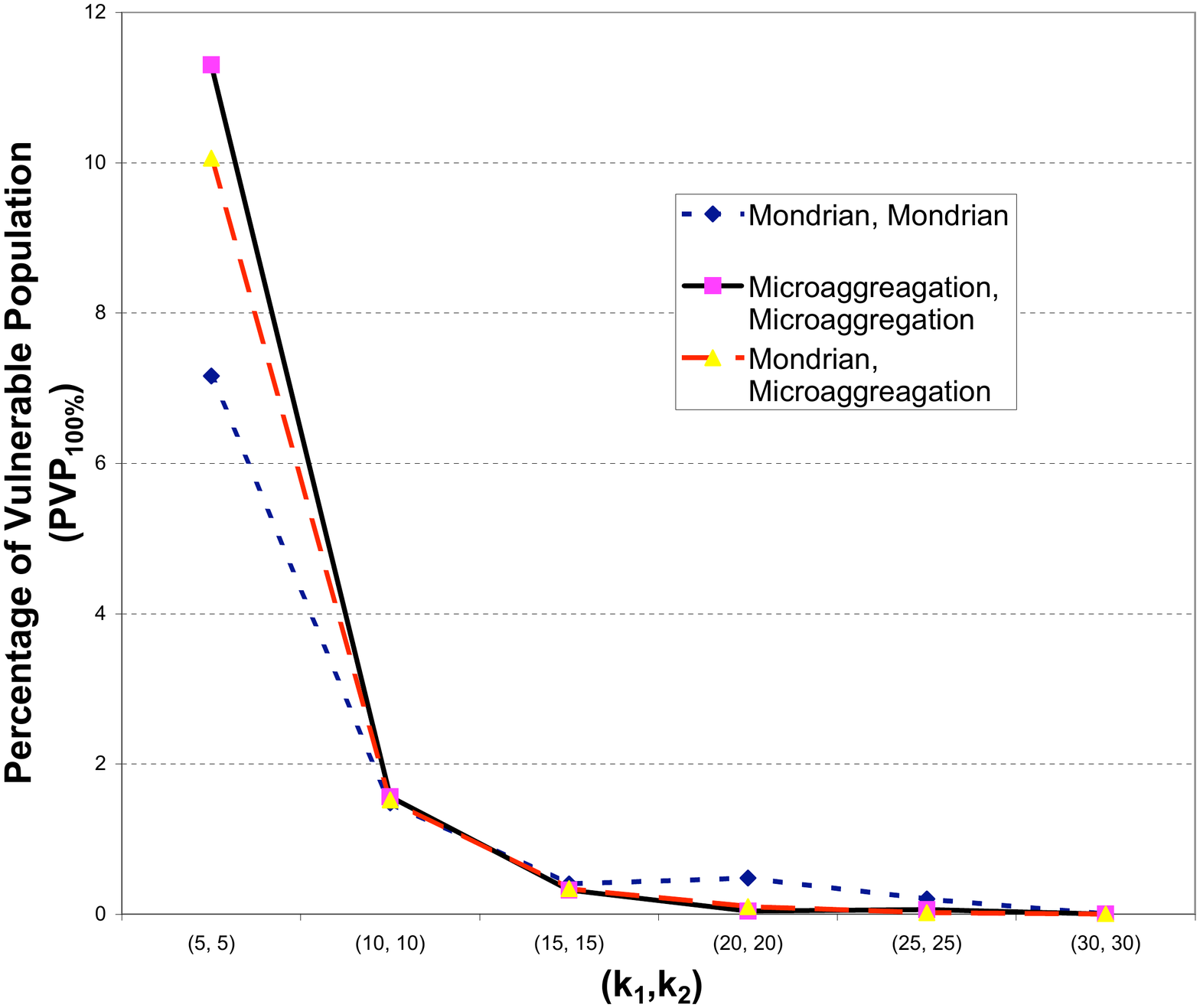}
&
\includegraphics[width=225pt, height=150pt]{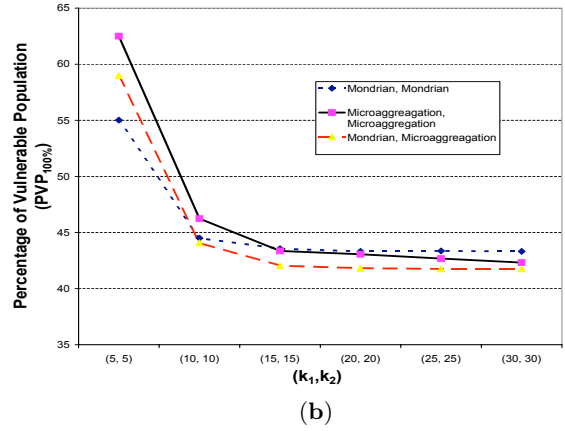} \\
 \hspace*{18pt}\bf(a) & \hspace*{32pt} \bf(b) 
\end{array}$
\end{footnotesize}
\caption{Severity of the intersection attack - perfect breach (a) Adult database (b) IPUMS database.}
\label{fig1}
\end{figure*}

A perfect breach occurs when the adversary can deduce the exact sensitive value of an individual. In other words, a perfect breach is when the adversary has a confidence level of 100\% about the individual's sensitive data. To estimate the probability of a perfect breach, we compute the percentage of overlapping population for whom the intersection attack leads to a final sensitive value set of size $1$. Figure~\ref{fig1} plots this result. %for the Adult and IPUMS databases.

We consider three scenarios for anonymizing the two overlapping subsets: (i) Mondrian on both the data subsets, %(Mondrian, Mondrian), 
(ii) Microaggregation on both the data subsets, and %(Microaggregation, Microaggregation), and 
(iii) Mondrian on the first subset and microaggregation on the second subset. %(Mondrian, Microaggregation). 
$(k_1,k_2)$ represents the pair of $k$ values used to anonymize the first and the second subset, respectively. In the experiments, we use the same $k$ values for both the subsets $(k_1 = k_2)$. %The $y$-axis represents the percentage of vulnerable population for whom the adversary has 100\% confidence level $(=\mathrm{\it PVP}_{1})$.
Note that for simplicity, from now on we will be defining confidence level in terms of percentages. %so $\mathrm{\it PVP}_{1}=\mathrm{\it PVP}_{100\%}$.   

In the case of Adult database we found that around 12\% of the population is vulnerable to a perfect breach for $k_1=k_2=5$. For the IPUMS database, this value is much more severe around 60\%. %It is clear to observe that the intersection attack works independent of the database considered and as such is not an artifact of a specific database. 
As the degree of anonymization increases or in other words, as the value of $k$ increases, the percentage of vulnerable population goes down. The reason for that is that as the value of $k$ increases, the partition sizes in each subset increases. This leads to a larger intersection set and thus lesser probability of obtaining an intersection set of size $1$.

\subsubsection{\hspace*{-6pt} Partial Breach}
\begin{figure*}[!htb]
\begin{footnotesize}
\centering $\begin{array}{cc}
\includegraphics[width=225pt, height=150pt]{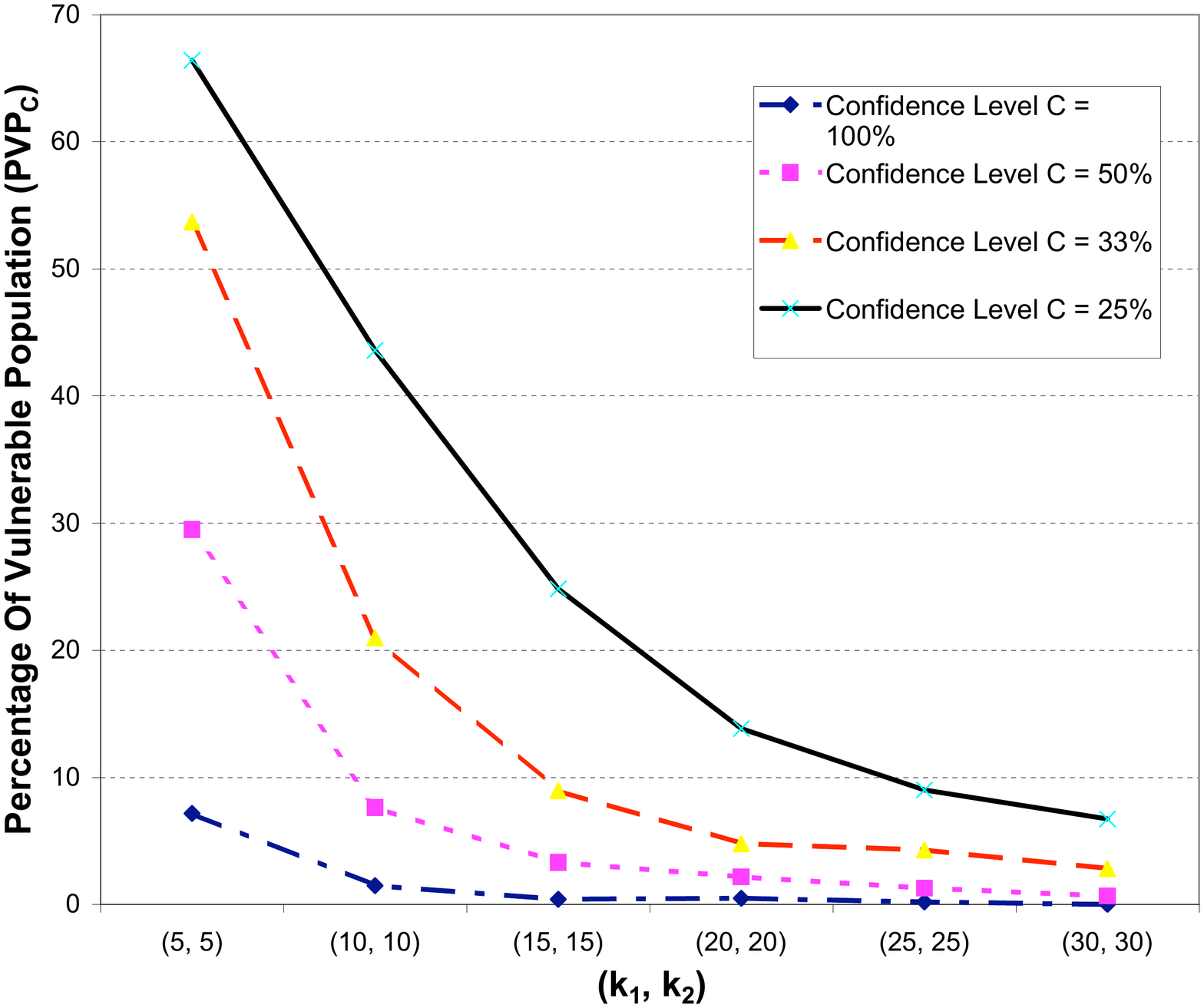}
&
\includegraphics[width=225pt, height=150pt]{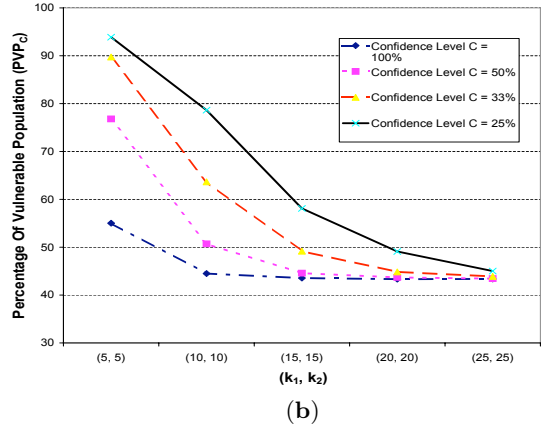} \\
\hspace*{18pt}\bf(a) & \hspace*{22pt} \bf(b) 
\end{array}$
\end{footnotesize}
\caption{Severity of the intersection attack - partial breach (a) Adult database (b) IPUMS database.}
\label{fig2}
\end{figure*}

Our next experiment aims to compute a more practical quantification of the severity of the intersection attack. In most cases, to inflict a privacy breach, all that the adversary needs to do is to boil down the possible sensitive values to a \, \emph{few} values which itself could reveal a lot of information. For example, for a hospital discharge database, by boiling down the sensitive values of the disease/diagn\-osis to a few values, say, ``Flu'', ``Fever'', or ``Cold'', it could be concluded that the individual is suffering from a viral infection. In this case, the adversary's confidence level is $1/3 = 33\%$. Figure~\ref{fig2} plots the percentage of vulnerable population for whom the intersection attack leads to a partial breach for the Adult and IPUMS databases. 

Here, we only use the first anonymization scenario described earlier in which both the overlapping subsets of the database are anonymized using Mondrian multidimensional technique. %Here again, the $x$-axis represents the pair of $k$ values used to anonymize the two subsets $(k_1, k_2)$. The $y$-axis represents the percentage of vulnerable population for whom the adversary had a confidence level of $C$ i.e. $(PVP_{C})$. 
Observe that the severity of the attack increases alarmingly for slight relaxation on the required confidence level.  For example, in the case of IPUMS database, around 95\% of the population was vulnerable for a confidence level of 25\% for $k_1=k_2=5$. For the Adult database, although this value is not as alarming, more than 60\% of the population was affected. %This indicates the practical quantification of privacy breach possible through the intersection attack. 

\subsection{Drop in Anonymity}

Our next goal is to measure the drop in anonymity occurring due to the intersection attack.% in partitioning-based anonymization schemes. 
To achieve this, we first take a closer look at the way these schemes work. As described in the earlier sections, the basic paradigm in partitioning-based anonymization schemes is to partition the data such that each partition size is at least $k$. The methodology behind partitioning and then summarizing varies from scheme to scheme. The minimum partition-size $(k)$ is thus used as a measure of the anonymity offered by these solutions.  However, the effective (or true) anonymity supported by these solutions is far less than the presumed anonymity $k$ (refer to the discussion in Section~\ref{sec:intattack}). % As discussed earlier in Section~\ref{sec:intattack} the effective anonymity offered can be computed by counting the number of distinct sensitive values present in the partition corresponding to each individual. The average effective anonymity is now the effective anonymity averaged over the entire overlapping population. 

%Although the aim is to achieve a partition size of at least $k$, in reality these schemes tend to produce partitions of size much larger than $k$. Furthermore,

\begin{figure*}[!htb]
\begin{footnotesize}
\centering $\begin{array}{cc}
\includegraphics[width=225pt, height=150pt]{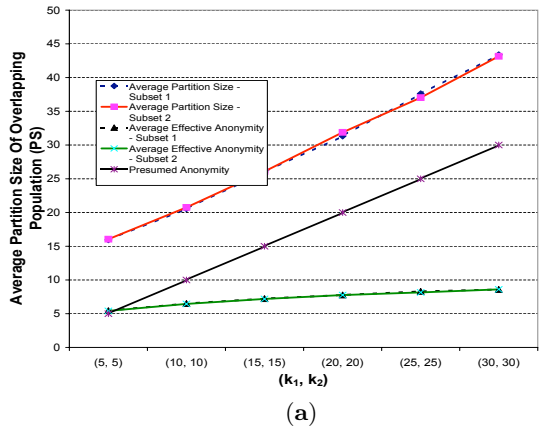}
&
\includegraphics[width=225pt, height=150pt]{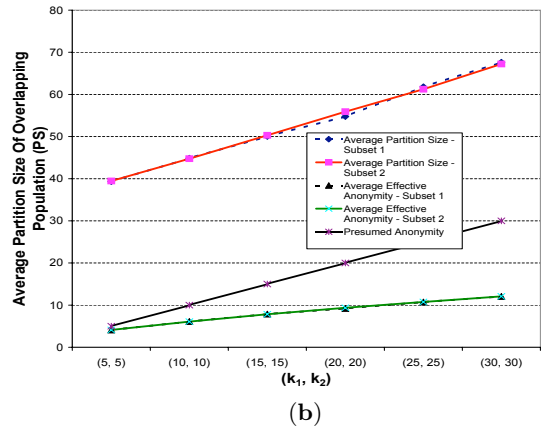} \\
\hspace*{23pt}\bf(a) & \hspace*{24pt} \bf(b) 
\end{array}$
\end{footnotesize}
\caption{Comparison of presumed anonymity, actual partition sizes, and effective anonymity \textbf(a) Adult database \textbf(b) IPUMS database.}
\label{fig3a}
\end{figure*}

Figure~\ref{fig3a} plots the average partition sizes and the average effective anonymities for the overlapping population. Here again, we only consider the scenario where both the overlapping subsets are anonymized using Mondrian multidimensional technique. Observe that the effective anonymity is much less than the partition size for both the data subsets. Also, note that these techniques result in partition sizes that are much larger than the minimum required of $k$. For example, the average partition size observed in the IPUMS database for $k=5$ is close to $40$. To satisfy the $k$-anonymity definition, there is no need for any partition to be larger than $2k+1$. The reasoning for this is straightforward as splitting the partition of size greater than $2k+1$ into two we get partitions of size at least $k$. Additionally, splitting any partition of size $2k+1$ or more only results in preserving more information. The culprit behind the larger average partition sizes is generalization based on user-defined hierarchies. Since generalization-based partitioning cannot be controlled at finer levels, the resulting partition sizes tend to be much larger than the minimum required value. 

%Coming back to the goal of measuring the drop in anonymity, we consider the effective anonymity as the measure of true anonymity offered by these anonymization techniques prior to intersection attack. So, 

\begin{figure*}[!htb]
\begin{footnotesize}
\centering $\begin{array}{cc}
\includegraphics[width=225pt, height=150pt]{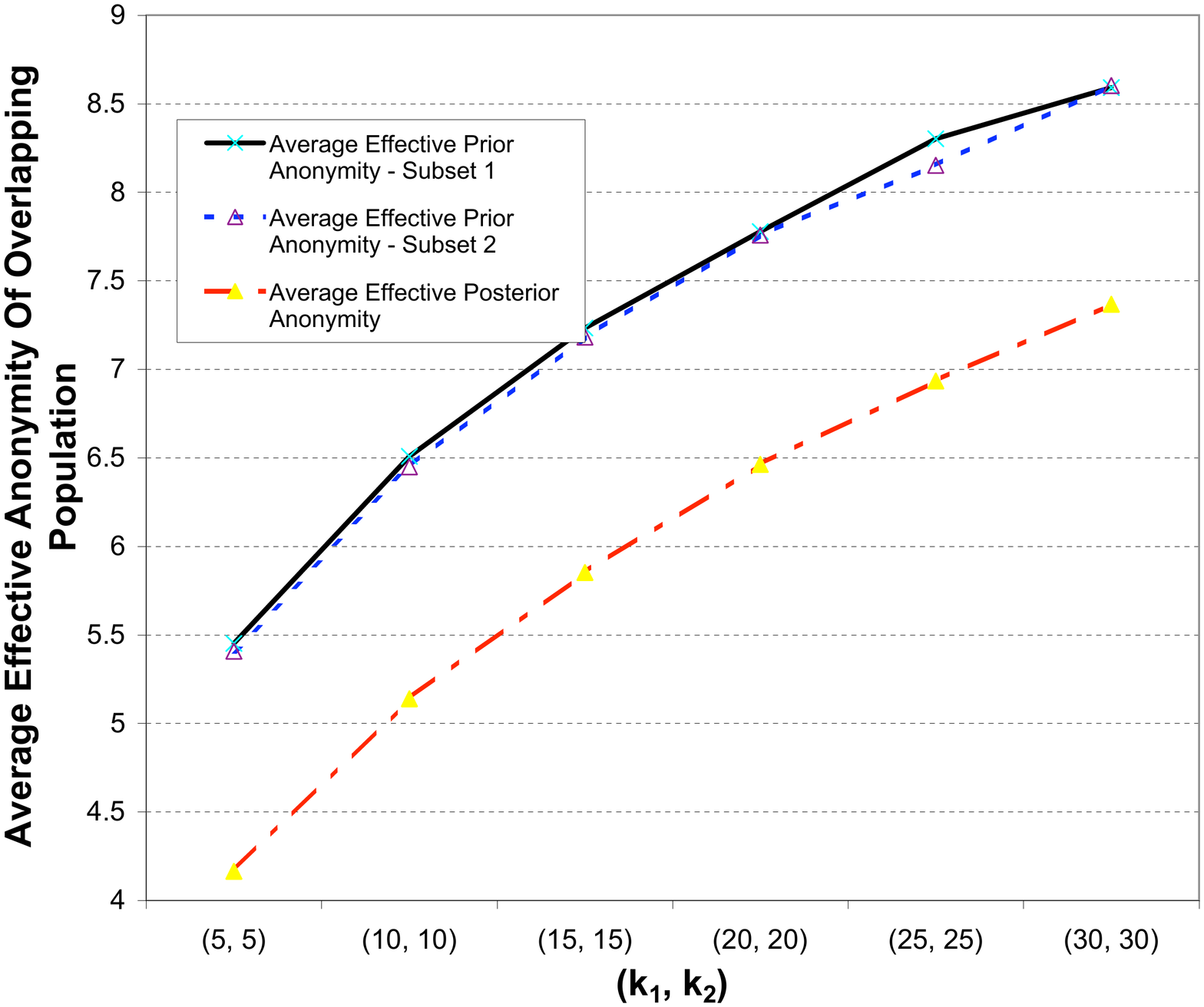}
&
\includegraphics[width=225pt, height=150pt]{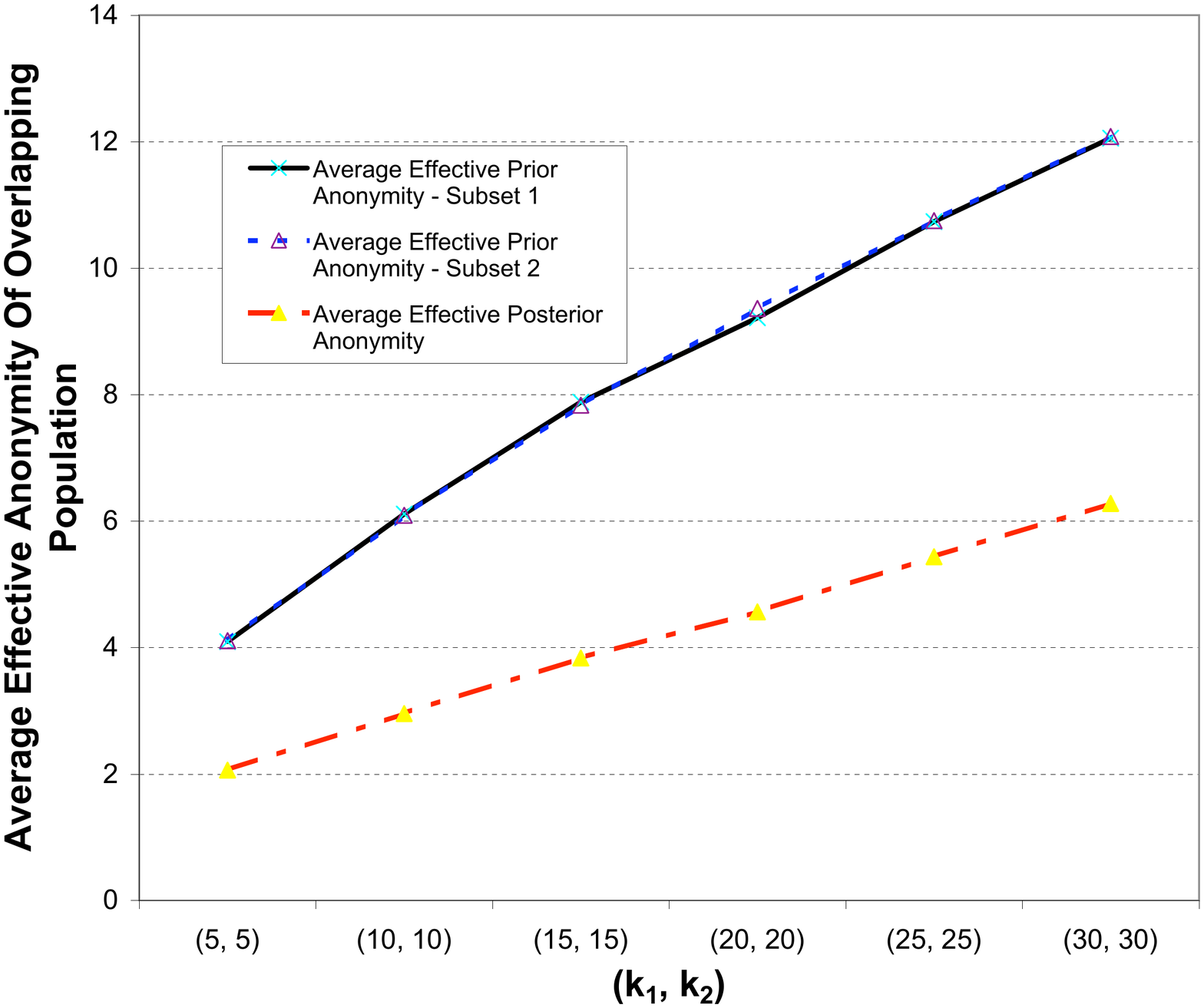} \\
\hspace*{23pt}\bf(a) & \hspace*{22pt} \bf(b) 
\end{array}$
\end{footnotesize}
\caption{Average drop in effective anonymity due to the intersection attack \textbf(a) Adult database \textbf(b) IPUMS database.}
\label{fig3b}
\end{figure*}

For each individual in the overlapping population, the effective prior anonymity is equal to the effective anonymity. 
We define the average effective prior anonymity with respect to a release as effective prior anonymities averaged over the individuals in the overlapping population. %We then measure the effective posterior anonymity after the intersection attack by counting the number of distinct sensitive values in the intersection of sensitive value sets retrieved from both the anonymized subsets. 
Similarly, the average effective posterior anonymity is the effective posterior anonymities averaged over the individuals in the overlapping population. The difference between the average effective prior anonymity and average effective posterior anonymity gives  the average drop in effective anonymity occurring due to the intersection attack. Figure~\ref{fig3b} plots the average effective prior anonymities and the average effective posterior anonymities for the overlapping population. Observe that the average effective posterior anonymity is much less than the average effective prior anonymity for both subsets. Also note that we measure drop in anonymities by using effective anonymities instead of presumed anonymities.  The situation only gets worse (drops get larger) when presumed anonymities are used.

 %after considering the true anonymity offered by the corresponding anonymization schemes vs.\ the presumed anonymity offered by these schemes. %(which is equal to the minimum partition size $k$ and was found to be much higher than the effective/true anonymity in the previous result).

\subsection{$\ell$-diversity and $t$-closeness}
\begin{figure*}[!htb]
\centering $\begin{array}{cc}
\includegraphics[width=225pt, height=150pt]{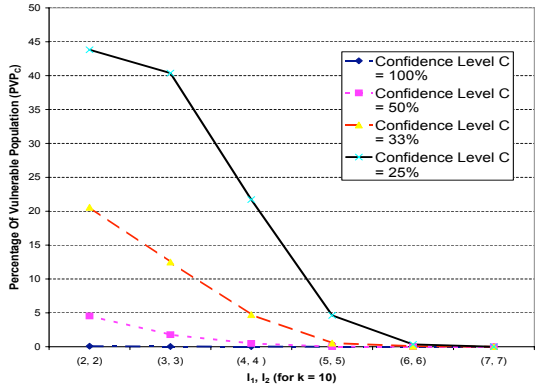}
&
\includegraphics[width=225pt, height=150pt]{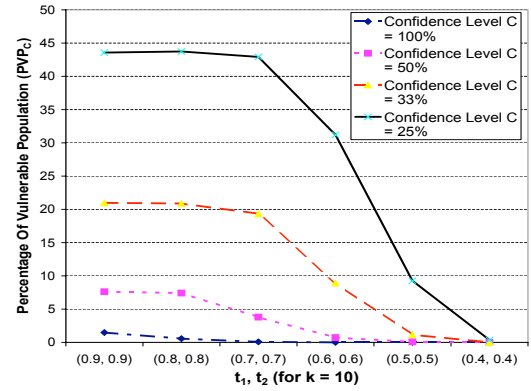} \\
\bf(a) l-diversity & \bf(b) t-closeness \\
\includegraphics[width=225pt, height=150pt]{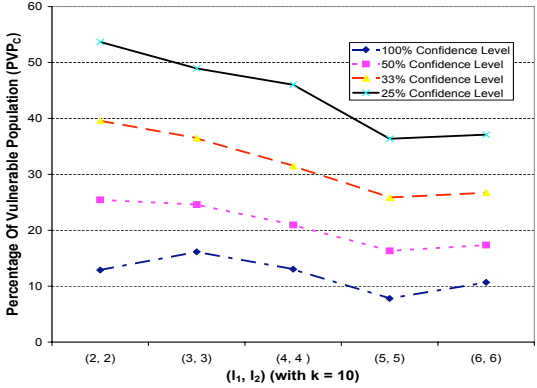}
&
\includegraphics[width=225pt, height=150pt]{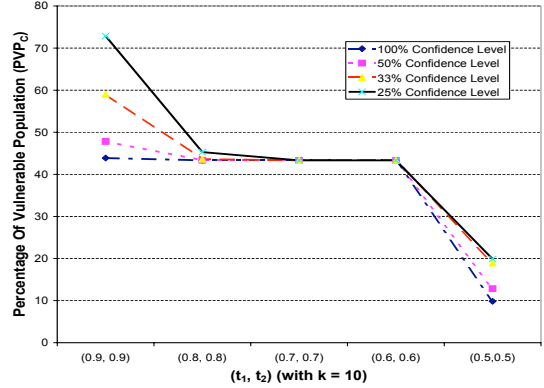} \\
\bf(c) l-diversity &(d) \bf t-closeness \\
\end{array}$
\caption{Severity of the intersection attack - l-diversity and t-closeness \textbf(a)(b) Adult Database \textbf(c)(d) IPUMS Database}
\label{fig4}
\end{figure*}

\begin{figure*}[!htb]
\centering $\begin{array}{cc}
\includegraphics[width=225pt, height=150pt]{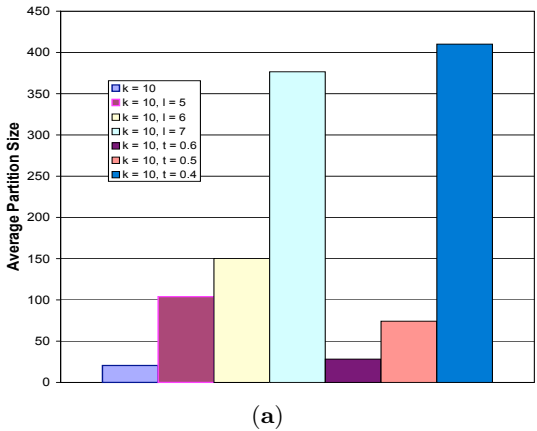}
&
\includegraphics[width=225pt, height=150pt]{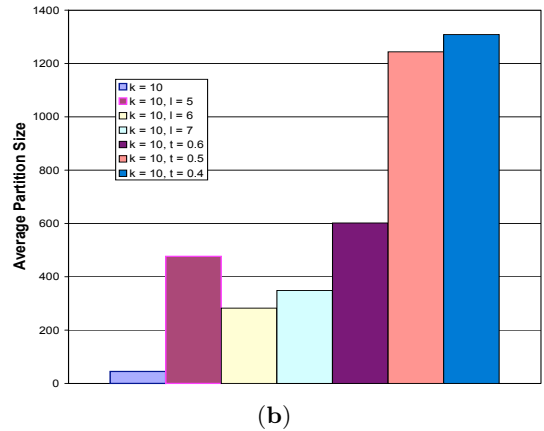} \\
\bf(a) & \bf(b) \\
\end{array}$
\caption{Average partition sizes for $\ell$-diversity and $t$-closeness \textbf(a) Adult Database \textbf(b) IPUMS Database}
\label{fig5}
\end{figure*}

We now consider the $\ell$-diversity and $t$-closeness extensions to the original $k$-anonymity definition. The goal again is to quantify the severity of the intersection attack by measuring the extent to which a partial breach occurs with varying levels of adversary confidence levels. Figure~\ref{fig4} plots the percentage of vulnerable population for whom the intersection attack leads to a partial breach for the Adult and IPUMS databases. Here, we anonymize both the subsets of the database with the same definition of privacy. We use the mondrian multidimensional $k$-anonymity with the additional constraints as defined by $\ell$-diversity and $t$-closeness. Figure~\ref{fig4}(a) plots the result for the $\ell$-diversity using the same $\ell$ value for both the subsets ($\ell_1 = \ell_2$) and with $k=10$. Figure~\ref{fig4}(b) plots the same for $t$-closeness. %using the same $t$ value for both the subsets ($t_1 = t_2$) with $k=10$. 
Even though these extended definitions seem to perform better than the original $k$-anonymity definition, they still lead to considerable breach in case of an intersection attack. This result is fairly intuitive in the case of $\ell$-diversity. Consider the definition of $\ell$-diversity: the sensitive value set corresponding to each partition should be ``well'' ($\ell$) diverse. However, there is no guarantee that the intersection of two well diverse sets leads to a well diverse set. $t$-closeness fares similarly. %and seems less vulnerable than $k$-anonymity. %However, this benefit comes at a cost. Both 
Also, both these definitions tend to force larger partition sizes, thus resulting in heavy information loss. Figure~\ref{fig5} plots the average partition sizes of the individuals corresponding to the overlapping population. It compares the partition sizes observed for $k$-anonymity, $\ell$-diversity, and $t$-closeness. For the IPUMS database, with a value of $k=10$, $k$-anonymity produces partitions with an average partition size of $45$. While, for the same value of $k=10$, with a value of $l=5$, the average partition size obtained was close to $450$. The partition sizes for $t$-closeness get even worse, where a combination of $k=10$ and $t=0.4$ yield partitions of average size close to $1300$. We can observe similar results for the Adult database.
%Further, we realized that the information loss due to these techniques is much higher in typical scenarios where the sensitive attribute domain sizes are large and discuss this further in the next section. 

%\begin{figure*}[!htb]
%\begin{footnotesize}
%\centering $\begin{array}{cc}
%\includegraphics[width=225pt, height=150pt]{fig4a}
%&
%\includegraphics[width=225pt, height=150pt]{fig4b} \\
%\hspace*{23pt}\bf(a) & \hspace*{27pt} \bf(b) 
%\end{array}$
%\end{footnotesize}
%\caption{Severity of the intersection attack for the IPUMS database (a) $\ell$-diversity (b) $t$-closeness.}
%\label{fig4}
%\end{figure*}

%\begin{figure}[!htb]
%\includegraphics[width=225pt, height=150pt]{fig5}
%\caption{Average partition sizes for $\ell$-diversity and $t$-closeness for the IPUMS database.}
%\label{fig5}
%\end{figure}

\subsection{Role of Sensitive Attribute Domain}
In all of the above experiments we use the ``Occupation'' (occupation code of the individual) as the sensitive attribute for both Adult and IPUMS databases as shown in Tables~\ref{adultdesc} and \ref{ipumsdesc}. The domain size of the Occupation attribute in the Adult database was $14$ whereas, the domain size in the IPUMS database was $247$. One of the plausible reasons for the attack to be more severe in case of the IPUMS database was the size of the sensitive attribute domain. This is because most of partition sizes are way larger than the minimum value required i.e. $k$, in case of the Adult database, it is possible that the sensitive value set corresponding to every partition contains all the possible values in the domain. This implies that an intersection of two sensitive value sets results in a set of size close to the size of the domain. Thus, it is possible that intersection attack will be less effective in cases where the sensitive attribute domain size is less than the average partition size. Intuitively, it seems like that in cases where the sensitive attribute domain size is large (of the order of several hundreds) the intersection attack would be more severe. Also, most real-life databases have sensitive attributes with large domain sizes. For example, if we consider a typical hospital discharge database, an ICD9 code is used to describe the diagnosis given to the patient. The possible values for this code is a number from $1$ to $999$~\cite{icd9} indicating the code for the specific patient diagnosis. In other cases, the sensitive attribute domain sizes tend be larger than this. The conjecture is that as the number of possible sensitive values increases, the intersection of two different sets results in a less diverse set. 

\begin{figure}[!htb]
\includegraphics[width=225pt, height=150pt]{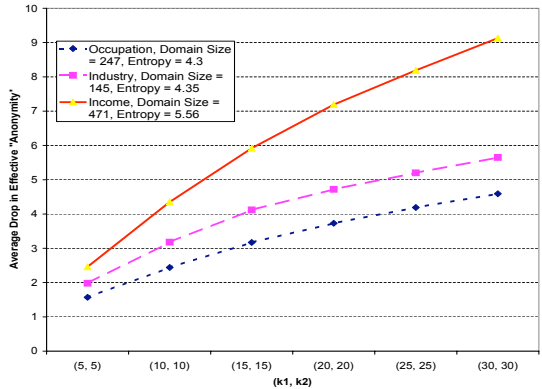}
\caption{Effect of sensitive attribute domain size - IPUMS database.}
\label{fig6}
\end{figure}

In order to confirm this, we constructed two new versions of the IPUMS database by replacing the sensitive attribute ``Occupation'' of each individual with ``Industry'' corresponding to the individual's work and ``Income'' corresponding to the total income of the individual. The domain sizes corresponding to these attributes is summarized in Table~\ref{ipumsversioning}. The domain size for ``Industry'' attribute is $145$, for the original ``Occupation'' attribute si $247$ and that of ``Income'' is $471$. Table~\ref{ipumsversioning} summarizes this. We ran the intersection attack on these new versions of the IPUMS database and compared it with the original. Figure~\ref{fig6} plots the average drop in effective anonymity for the overlapping population. Based on our conjecture, the drop in effective anonymity should increase with the increase in the sensitive attribute domain size. Surprisingly we did not observe the trend we were expecting. The drop in effective anonymity in case of ``Occupation'' was less than when compared with ``Industry''. It turns out that the reason for this is that the \emph{actual} number of possible values for each sensitive attribute does not necessarily be the same as the domain size, or in other words the \emph{total} number of possible values. So, a large sensitive attribute domain size does not guarantee that the number of possible values actually occuring is large. Instead, a simple entropy measure such as the shannon's entropy could be used to measure the actual number of possible values. The entropy value for each of these attributes is listed in Table~\ref{ipumsversioning}. Although the actual domain size for `Occupation'' attribute is larger, its entropy is less than that of than that of the ``Industry'' attribute. Now, the conjecture is that as the entropy (or information content) of the sensitive attribute increases, the severity of intersection attack increases. Our result in Figure~\ref{fig6} confirms this. The average drop in effective anonymity increases with the entropy of the corresponding sensitive attribute domain since the non-sensitive attributes are kept the same for all the datasets.

%So, as the number of possible sensitive values increases, the intersection of two different sets results in a less diverse set. However, a large sensitive attribute domain size does not guarantee that the number of possible values actually occuring is large. A simple diversity measure such as the shannon's diversity measure could be used instead. The conjecture is that as the diversity of the sensitive attribute increases, the severity of intersection attack increases. In order to confirm this, we constructed two new versions of the IPUMS database by replacing the sensitive attribute ``Occupation'' of each individual with ``Industry'' corresponding to the individual's work and ``Income'' corresponding to the total income of the individual. The domain sizes and diversities corresponding to these attributes and the original sensitive attribute ``Occupation'' is listed in \ref{ipumsversions}. These poverty attribute had a domain size of ... and the income attribute had a domain size of ... Table~\ref{ipumsversioning} summarizes this. We ran the intersection attack on these new versions of the IPUMS database and compared it with the original. Figure~\ref{fig6} plots the average drop in effective anonymity for the overlapping population. Observe that the average drop in effective anonymity increases as the sensitive attribute diversity increases. 

\begin{table}
	\centering
		\begin{tabular}{||c|c|c||} \hline
			\textbf{Sensitive Attribute}	& \textbf{Domain Size}	& \textbf{Diversity} \\ \hline
			Occupation	&247 &4.30 \\ \hline
			Industry	&145	&4.35 \\ \hline
			Income	&471	&5.56 \\ \hline
		\end{tabular}	
		\caption{IPUMS database versions (Non-Sensitive attributes remain same as the original)}
	\label{ipumsversioning}
\end{table}

\subsection{Number of Databases}
\begin{figure*}[!htb]
\centering $\begin{array}{cc}
\includegraphics[width=225pt, height=150pt]{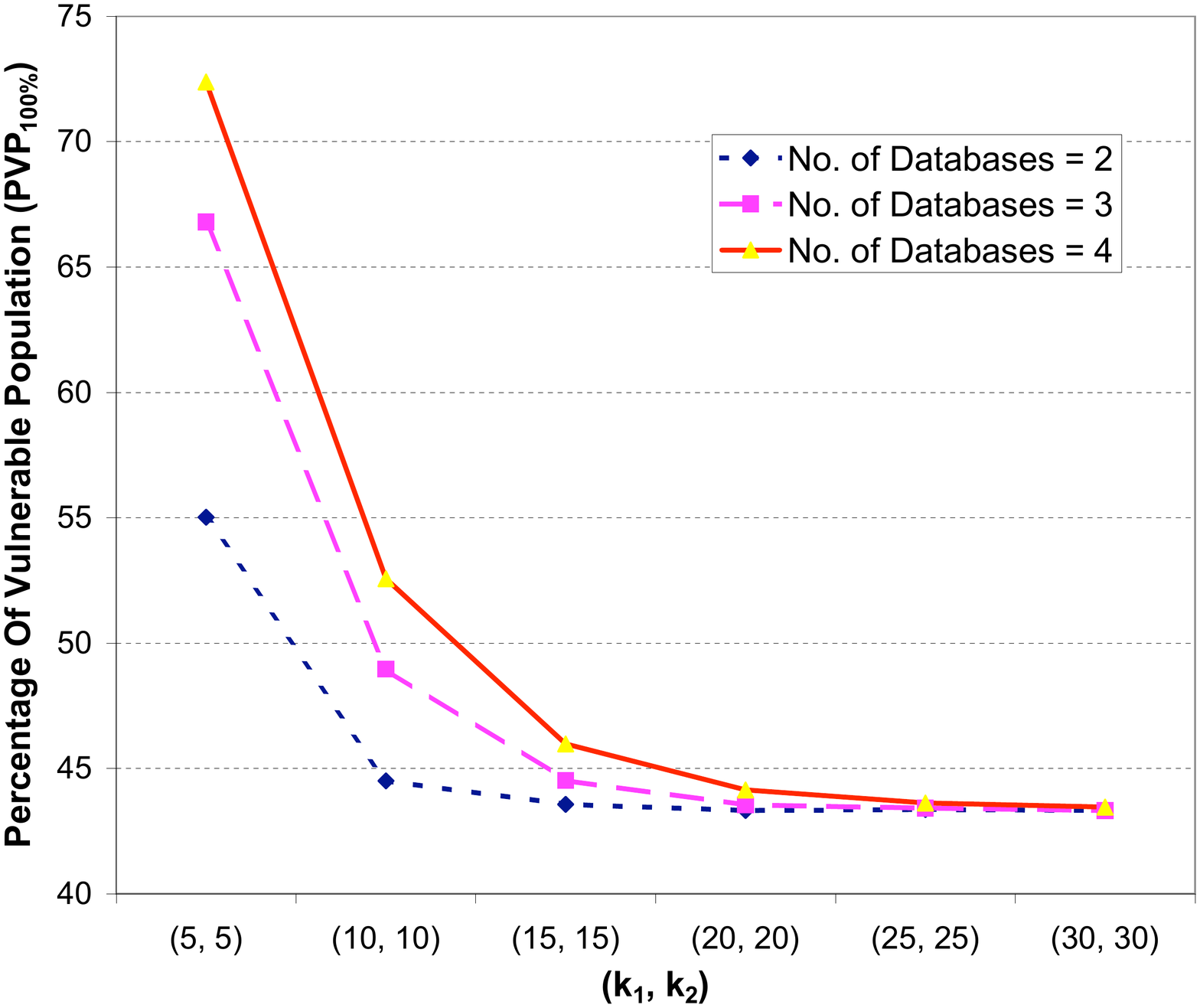}
&
\includegraphics[width=225pt, height=150pt]{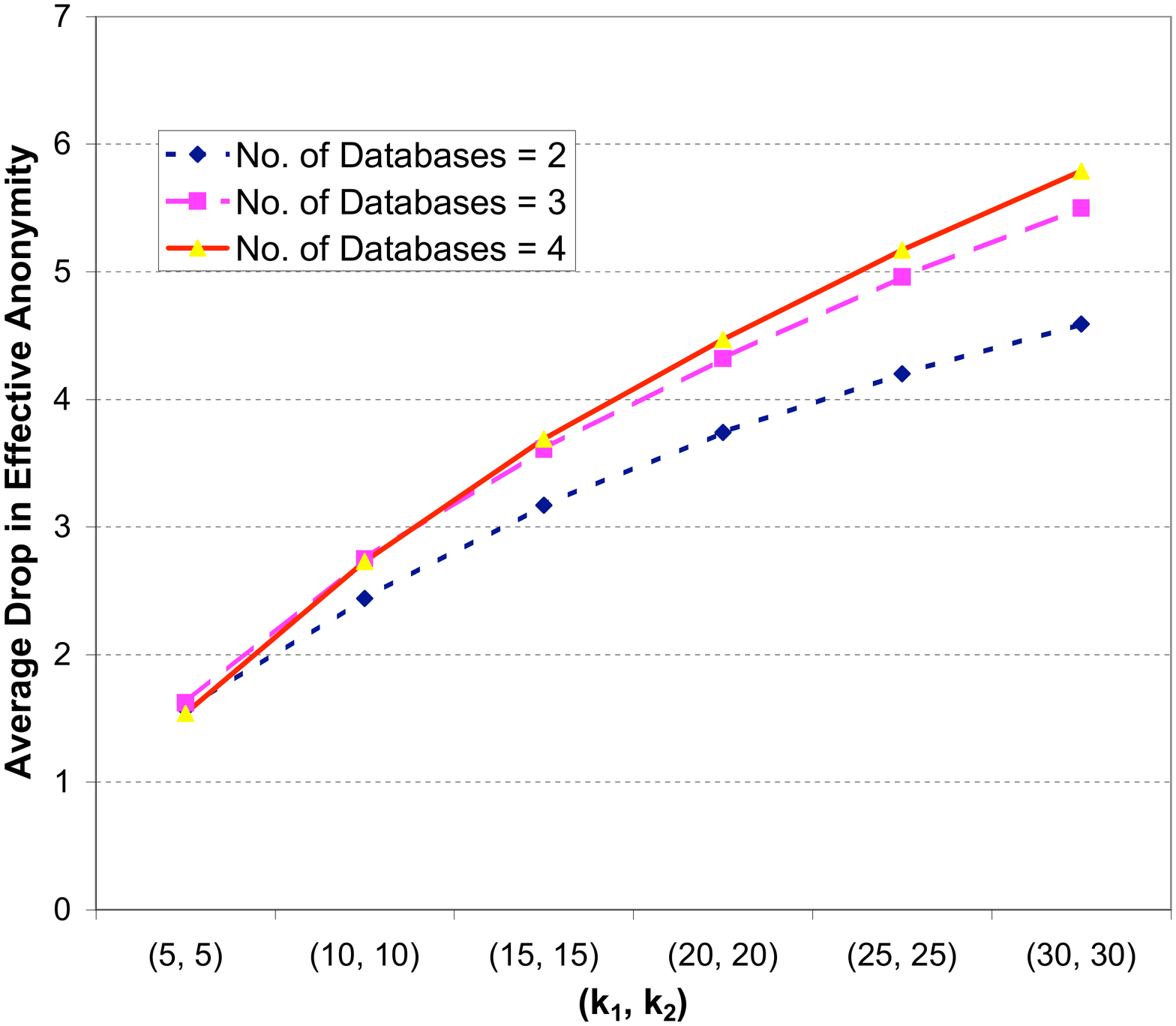} \\
\bf(a) & \bf(b) \\
\end{array}$
\caption{Effect of Number of Anonymized Releases - IPUMS Database \textbf(a) Percentage of Vulnerable Poplation \textbf(b) Drop in Effective Anonymity}
\label{fig7}
\end{figure*}

In the above experiments we have considered the scenario in which two anonymized releases contain information about overlapping population. As data publishing becomes more prevelant among organizations that would like to share data for research and collaborative purposes, it is possible that the number of anonymized releases available containing information about the same subset of people is more than just two. The adversary could use as many anonymized releases as possible to gather information about a target population and use the intersection attack to deduce the sensitive attribute values. In such a scenario, it is interesting to see how the intersection attack performs in the presence multiple (more than 2) overlapping anonymized releases. We first consider the percentage of vulnerable population with a confidence level of $100\%$ $(PVP_{100\%})$. Figure~\ref{fig7}(a) plots this for varying number $(n = 2,3,4)$ of anonymized releases available to adversary. Here again, we build $n$ overlapping subsets of the IPUMS database by fixing the overlapping population at $5000$. It can be observed that the severity of the intersection attack increases with the increase in the number of anonymized releases available to the adversary. There is a significant increase in the percentage of vulnerable population with the increase in $n$, for small values of $k$. However, there seem to be no such significant increase for larger values of $k$. The reason for this is that the partition sizes for larger values of $k$ tend to be large enough such that the presence of additional anonymized releases does not help the intersection attack anymore. Alternative to the severity of the attack, we can study the effect of the number of anonymized releases on the drop in effective anonymity. Figure~\ref{fig7}(b) plots the average drop in effective anonymity for varying number $(n = 2,3,4)$ of anonymized releases. Here again we can observe that drop in effective anonymity increases with the increase in the number of anonymized releases. These results indicate that if the anonymized releases correspond to fairly larger values of $k$, there is only limited information gained by the adversary by collecting additional releases.  

%\subsection{Choice of Algorithm}

\remove{\subsection{Experimental Conclusions}
From the above results, we conclude the~following:
\begin{CompactItemize}
\item Existing partition-based anonymization approaches are vu\-lnerable to composition attacks. These methods seem to mitigate composition attacks to some extent by producing artificially large clusters. Refining the algorithms to produce finer clusters would not help as it will only increase the severity of the intersection attack.
\item The extended definitions of $\ell$-diversity and $t$-closeness fare better than the original $k$-anonymity definition but still lead to considerable breach in the case of an intersection attack. Additionally, these schemes lead to huge partition sizes and thus in heavy information loss.
\item Real-life databases typically have sensitive attributes with large domain sizes and our experimental results indicate that the severity of the intersection attack increases with the sensitive attribute domain size.
\end{CompactItemize}
}

%\documentclass[10pt,twocolumn]{article}

%\input{corr-preamble}

%%
%\begin{document}

\section{Differential Privacy} \label{sec:diffprivacy}

In this section we give a precise formulation of ``resistance to arbitrary side information'' and show that several relaxations of differential privacy imply it.  The formulation follows the ideas originally due to Dwork and McSherry, stated implicitly in \cite{Dwork06}. This is, to our knowledge, the first place such a formulation appears explicitly. The proof that relaxed definitions (and hence the schemes of \cite{DKMMN06,NRS07,MKAGV08}) satisfy the Bayesian formulation is new. These results are explained in a greater detail in a separate technical report~\cite{KS08}. In this paper we just reproduce the relevant parts from~\cite{KS08}.

We represent databases as vectors in $\D^n$ for some domain $\D$ (for example, in the case of the relational databases above, $\D$ is the product of the attribute domains). There is no distinction between ``sensitive'' and ``insensitive'' information. Given a randomized algorithm $\AAA$, we let $\AAA(D)$ be the random variable (or, probability distribution on outputs) corresponding to input $D$.  

\begin{definition}[Differential Privacy] 
\label{def:ind}
A randomized algorithm $\sampnum$ is \mbox{\bf \boldmath $\eps$-differentially} {\bf private} if for all databases $D_1,D_2\linebreak \in \D^n$ that differ in one individual, and for all subsets $S$ of outputs, \linebreak $\Pr[\sampnum (D_1) \in S] \leq e^{\eps} \Pr[\sampnum (D_2)\in S].$
\end{definition}

This definition states that changing a single individual's data in the database leads to a small change in the {\em distribution} on outputs. Unlike more standard measures of distance such as total variation (also called statistical difference) or Kullback-Leibler divergence, the metric here is multiplicative and so even very unlikely events must have approximately the same probability under the distributions $\AAA(D_1)$ and $\AAA(D_2)$. This condition  was relaxed somewhat in other papers \cite{DiNi03,DwNi04,BDMN05,DKMMN06,CM06,NRS07,MKAGV08}. The schemes in all those papers, however,  satisfy the following relaxation~\cite{DKMMN06}:

\begin{definition}%[$(\eps,\delta)$-Diff.\,Privacy] 
\label{def:indd}
A randomized algorithm $\sampnum$ is {\bf \boldmath $(\eps,\delta)$-dif\-fer\-en\-tially private} if for all databases $D_1,D_2\in \D^n$ that differ in one individual, and for all subsets $S$ of outputs, $\Pr[\sampnum (D_1)\in S] \leq e^{\eps} \Pr[\sampnum (D_2)\in S]+\delta\,.$
\end{definition}

The relaxations used in \cite{DwNi04,BDMN05,MKAGV08} were in fact stronger (i.e., less relaxed) than \defref{ind}. One consequence of the results below is that all the definitions are equivalent up to polynomial changes in the parameters, and so given the space constraints we work only with the simplest notion.\footnote{That said, some of the other relaxations, such as probabilistic differential privacy of ~\cite{MKAGV08}, could lead to better parameters in \thmref{ind2sdp}.}

\subsection{Semantics of Differential Privacy} \label{sec:bayes}

There is a crisp, semantically-flavored interpretation of differential privacy, due to Dwork and McSherry, and explained in \cite{Dwork06}: {\em Regardless of external knowledge, an adversary with access to the sanitized database draws the same conclusions whether or not my data is included in the original data.} (the use of the term ``semantic'' for such definitions dates back to semantic security of encryption \cite{GM84}).

We require a mathematical formulation of ``arbitrary external knowledge'', and of ``drawing conclusions''. The first is captured via a {\em prior} probability distribution $b$ on $\D^n$ ($b$ is a mnemonic for ``beliefs''). Conclusions are modeled by the corresponding posterior distribution: given a transcript $t$, the adversary updates his belief about the database $D$ using Bayes' rule to obtain a posterior~$\widehat{b}$:
\begin{equation} \label{eqn:bel}
\widehat{b}[D| t] \defeq \frac{\Pr[\sampnum (D)=t] b[D]}{\sum_{D'} \Pr[\sampnum (D')=t]b[D']}\ . 
\end{equation}
%This posterior distribution models the adversary's conclusions based on the output $t$. 
In an interactive scheme, the definition of $\sampnum$ depends on the adversary's choices; for simplicity we omit the dependence on the adversary in the notation. Also, for simplicity, we discuss only discrete probability distributions. Our results extend directly to the interactive, continuous case.
%here abuse notation somewhat and use $b[\cdot]$ and $\Pr[\cdot]$ to denote probability mass functions, for discrete distributions, and probability density functions for continuous distributions. This is done simply to avoid tangling with measure-theoretic issues.

For a database $D$, define $D_{-i}$ to be the vector obtained by replacing position $i$  by some default value in $D$ (any value in $D$ will do). This corresponds to ``removing'' person $i$'s data. We consider $n+1$ related scenarios (``games'', in the language of cryptography), numbered 0 through $n$. In Game 0, the adversary interacts with $\sampnum(D)$. This is the interaction that takes place in the real world. In Game $i$ (for $1\leq i \leq n$), the adversary interacts with $\sampnum(D_{-i})$. Game $i$ describes the hypothetical scenario where person $i$'s data is not included. 

For a particular belief distribution $b$ and transcript $t$, we consider the $n+1$ corresponding posterior distributions $\widehat{b}_0,\dots,\widehat{b}_n$. The posterior $\widehat{b}_0$ is the same as $\widehat{b}$ (defined in Eq.\,\eqref{eqn:bel}). For larger $i$, the $i$-th posterior distribution $\widehat{b}_i$ represents the conclusions drawn in Game $i$, that is
$$\widehat{b}_i[D| t] \defeq \frac{\Pr[\sampnum (D_{-i})=t] b[D]}{\sum_{D'} \Pr[\sampnum(D'_{-i})=t]b[D']}\,. $$

Given a particular transcript $t$,  privacy has been breached if there exists an index $i$ such that the adversary would draw different conclusions depending on whether or not $i$'s data was used. It turns out that the exact measure of ``different'' here does not matter much. We chose the weakest notion that applies, namely statistical difference. If $\erert$ and $\Q$ are probability measures on the set $\mathcal{X}$, the statistical difference between $\erert$ and $\Q$ is defined as:
$$\sd{\erert,\Q}= \max_{S \subset \mathcal{X}}|\erert[S]-\Q[S]|.$$
%We say there is a problem for transcript $t$ if the distributions $\widehat{b}_0[\cdot|t]$ and $\widehat{b}_i[\cdot|t]$ are far apart in statistical difference. We would like to avoid this happening for any potential participant. This is captured by the following definition.
\begin{definition}%[$\eps$-semantic privacy]  
\label{def:sem}
An algorithm $\sampnum$ is {\bf \boldmath $\eps$-semantically private} if for all prior distributions $b$ on $\D^n$, for all databases $D \in \D^n$, for all possible transcripts $t$, and for all $i = 1,\ldots,n$,
$$\sd{\widehat{b}_0[D|t]\ ,\  \widehat{b}_i[D|t]\ } \leq \eps.$$
%A possible transcripts means transcripts which occur with non-zero probability for some input, i.e., $\exists \widehat{D} \in \D^n, \Pr[t \in \AAA(\widehat{D})] > 0$. 
\end{definition}

\noindent This can be relaxed to allow a probability $\delta$ of failure.

\begin{definition}
An algorithm is {\bf \boldmath $ (\eps,\delta)$-semantically private} if, for all prior distributions $b$,  with probability at least $1-\delta$ over pairs $(D,t)$, where the database $D \from b$ $(D$ is drawn according to $b)$ and the transcript $t \from \AAA(D)$ $(t$ is drawn according to $\AAA(D))$, for all $i =1,\dots,n$: 
$\sd{\widehat{b}_0[D|t]\,,\,\widehat{b}_i[D|t]} \leq \eps. $
\end{definition}

Dwork and McSherry proposed the notion of semantic privacy, informally, and observed that it is equivalent to differential privacy.

%We now show that the notions of $\eps$-indistinguishability (Definition \ref{def:ind}) and $\eps$-semantic privacy (Definition \ref{def:sem}) are very closely related. This proves the fact that $\eps$-indistinguishability resists attacks from Bayesian adversaries, even when the adversaries have arbitrary side information. 

\begin{proposition}[Dwork-McSherry] \label{thm:eind}
$\eps$-differential privacy implies $\widehat\eps$-semantic privacy, where $\widehat\eps=e^{\eps}-1$. %Conversely, $\widehat\eps/2$-semantic privacy implies $2\eps$-differential privacy. 
\end{proposition}

\noindent
We show that this implication holds much more generally:
%The main technical result of this section is that this extends to the relaxations of~\cite{DKMMN06,NRS07,MKAGV08}.

\begin{theorem}[Main Result]\label{thm:ind2sdp}
($\eps,\delta$)-differential privacy implies $(\eps',\delta')$-semantic privacy where $\eps'=e^{3\eps}-1+2\sqrt{\delta}$ and $\delta' = O(n\sqrt{\delta})$. %Conversely, $(\widehat\eps/2,\delta)$-semantic privacy implies $(2\eps,2\delta)$-differential privacy.
\end{theorem}

\thmref{ind2sdp} states that the relaxations notions of differential privacy used in some previous work still imply privacy in the face of arbitrary side information. This is {\em not} the case for {\em all} possible relaxations, even very natural ones. For example, if one replaced the multiplicative notion of distance used in differential privacy with total variation distance, then the following ``sanitizer'' would be deemed private: choose an index $i\in\{1,\dots,n\}$ uniformly at random and publish the entire record of individual $i$ together with his or her identity (example 2 in \cite{DMNS06}). Such  a ``sanitizer'' would not be meaningful at all, regardless of side information.

\remove{Theorem~\ref{thm:ind2sdp} also gives some qualitative improvements over the privacy claims made in some previous work. For example, it %\thmref{ind2sdp-inf} 
implies that the claims of \cite{DwNi04,BDMN05} can be strengthened to hold simultaneously for \emph{all} predicates of the input (a switch in the order of quantifiers). }
%The strengthening does come at some loss in parameters since $\delta$ is increased. This incurs a factor of 2 in $\logdel$, or a factor of $\sqrt{2}$ in the standard deviation. 
%More significantly, \thmref{ind2sdp} shows that noise processes with negligible probability of bad events have nice differential privacy guarantees even for adversaries who are not necessarily informed. There is a hitch, however: only adversaries whose beliefs somehow represent reality, i.e. for whom the real database is somehow ``representative" of the adversary's view, can be said to learn nothing. 

Finally, the techniques used to prove \thmref{ind2sdp} can also be used to analyze schemes which do not provide privacy for {\em all} pairs of neighboring databases $D_1$ and $D_2$, but rather only for {\em most} such pairs (neighboring databases are the ones that differ in one individual). Specifically, it is sufficient that those databases where the ``indistinguishability'' condition fails occur with small probability.  

\begin{definition} [$(\eps,\delta)$-indistinguishability]
Two rand\-om variables $X,Y$ taking values in a set $\mathcal{X}$ are \textbf{$(\eps,\delta)$-indistingui\-shable} if for all sets $S\subseteq \mathcal{X}$, $\Pr[X\in S] \leq e^{\displaystyle \eps} \Pr[Y\in S] + \delta $ and $\Pr[Y\in S] \leq e^{\displaystyle \eps} \Pr[X\in S] + \delta\,.$
\end{definition}

\begin{theorem} \label{prop:dsemantic}
Let $\AAA$ be a randomized algorithm. Let $\mathcal{E}=\allowbreak\{D_1 \in \D^n: \allowbreak\forall \mbox{ neighbors }D_2 \mbox{ of } D_1,\ \allowbreak \AAA(D_1) \mbox{ and } \allowbreak\AAA(D_2)\ \allowbreak \mbox{are}\ \allowbreak \linebreak (\eps,\delta)\mbox{-indistinguishable}\}\,.$ Then $\AAA$ satisfies $(\eps',\delta')$-semantic privacy for any prior distribution $b$ such that $b[\mathcal{E}] = \Pr_{D_3 \from b}[D_3 \in \mathcal{E}] \geq 1-\delta$ with $\eps'=e^{3\eps}-1+2\sqrt{\delta}$ and $\delta'=O(n\sqrt{\delta})$. 
\end{theorem}

%\subsection{Proof Sketch for Main Results} Due to space constraints, we defer the proofs of Theorems \ref{thm:ind2sdp} and~\ref{prop:dsemantic} to the full version of this paper~\cite{arXiv:0803.0032}.  

\subsection{Proof Sketch for Main Results} The complete proofs are described in~\cite{KS08}. Here we sketch the main ideas behind both the proofs. Let $Y|_{X=a}$ denote the conditional distribution of $Y$ given that $X=a$ for jointly distributed random variables $X$ and $Y$. The following lemma (proof omitted) plays an important role in our proofs.

\begin{lemma}[Main Lemma]\label{lem:retro}
Suppose two pairs of random variables $(X,\sampnum(X))$ and $(Y,\sampnum'(Y))$ are $(\eps,\delta)$-indistinguishable (for some randomized algorithms $\AAA$ and $\AAA'$). Then with probability at least $1-\delta''$ over $t\from \sampnum(X)$ (equivalently $t \from \A'(Y)$), the random variables $X|_{\sampnum(X) =t}$ and $Y|_{\sampnum'(Y)=t}$ are $(\hat{\eps},\hat{\delta})$-indistinguish\-able with $\hat{\eps}=3\eps$, $\hat{\delta}=2\sqrt{\delta}$, and $\delta''=\sqrt{\delta}+\frac{2\delta}{\eps e^{\eps}}=O(\sqrt{\delta})$. 
%In particular, with probability at least $1-\delta''$ over $t\from \sampnum(X)$ $($equivalently $t \from \A'(Y)$$)$, the statistical difference between \linebreak $X|_{\sampnum(X) =t}$ and $Y|_{\sampnum'(Y)=t}$ is at most $\eps'=e^{\hat{\eps}}-1+\hat{\delta}$. 
\end{lemma}

Let $\AAA$ be a randomized algorithm (in the setting of Theorem~\ref{thm:ind2sdp}, $\AAA$ is a $(\eps,\delta)$-differentially private algorithm). Let $b$ be a belief distribution (in the setting of Proposition~\ref{prop:dsemantic}, $b$ is a belief with $b(\mathcal{E}) \geq 1-\delta$). The main idea behind both the proofs is to use Lemma~\ref{lem:retro} to show that with probability at least $1-O(\sqrt{\delta})$ over pairs $(D,t)$ where $D \from b$ and $t \from \AAA(D)$, $\sd{b|_{\sampnum(D) =t},b|_{\sampnum(D_{-i}) =t}} \leq \eps'$. Taking a union bound over all coordinates $i$, implies that with probability at least $1-O(n\sqrt{\delta})$ over pairs $(D,t)$ where $D \from b$ and $t \from \AAA(D)$, for all $i=1,\dots,n$, we have $\sd{b|_{\sampnum(D) =t},b|_{\sampnum(D_{-i}) =t}}$ $\leq  \eps'$. For Proposition~\ref{prop:dsemantic}, it shows that $\AAA$ satisfies $(\eps',\delta')$-semantic privacy for $b$. In the Theorem~\ref{thm:ind2sdp} setting where $\AAA$ is $(\eps,\delta)$-different\-ially private and $b$ is arbitrary, it shows that $(\eps,\delta)$-differential privacy implies $(\eps',\delta')$-semantic privacy.

%[Shiva, could you write something here? you like to fill this in? Keep it to about 1/2 of one column.]

%\footnotesize
%\bibliographystyle{abbrv}
%\bibliography{privacy-corr}  % sigproc.bib is the name of the Bibliography in this case

%\end{document}

\section{Concluding Remarks} \label{sec:conclusion}
In this paper we explored how one can reason about privacy in the presence of independent anonymized releases of overlapping population. %face of rich, realistic sources of auxiliary information.  %Specifically, we investigate the effectiveness of current anonymization schemes in preserving privacy when multiple organizations {\em independently} release anonymized data about overlapping populations. 
Our experimental study indicates that several currently proposed partition-based anonymization schemes, including $k$-anonymity and its variants, are vulnerable to composition attacks.  %For two different implementations of $k$-anonymity we found that sensitive information of a significant percentage of population could be compromised. %These methods seemed to somewhat mitigate these attacks by producing artificially large clusters. Refining the algorithms to produce finer clusters will only increase the severity of the intersection attack. 
%The $\ell$-diversity and $t$-closeness extensions fare better but still lead to considerable breach in the case of an intersection attack.  Additionally, these schemes lead to huge partition sizes, thus resulting in heavy information loss. 
On the positive side, we gave a precise formulation of the property ``resistance to arbitrary side information'' and show that several relaxations of differential privacy satisfy it. %This allows for modular use of differentially-private schemes. 

The most striking question that arises from this work is whether randomness in the anonymization algorithm is {\em necessary} to resist complex side information such as independent releases. Another interesting direction would be to study other settings where composition attacks are realistic and effective? A natural candidate for future investigation are the releases of overlapping contingency tables that are often considered in the statistical literature.

% If so, are there important types of functionality that are provided by deterministic algorithms and which provably cannot be provided by differentially private mechanisms? 

\remove{More generally, in what other settings are composition attacks realistic and effective? A natural candidate for future investigation are the releases of overlapping contingency tables that are often considered in the statistics literature.  Is it possible to find a taxonomy of basic attacks (linkage attacks, composition attacks, link analysis in social networks, etc) that would inform the development of new anonymization schemes, much the way modern cryptanalysis informs the design of new signature and encryption schemes? What should be the main pieces of such a taxonomy?}

\begin{small}

\bibliographystyle{abbrv}
\bibliography{../bibfiles/master}  % sigproc.bib is the name of the Bibliography in this case

\end{small}

%\section{appendix1}
%
%\begin{enumerate}
%
%\item Grouping based methods seem inherently vulnerable to composition attacks.
%
%\item Existing grouping based methods are vulnerable to a simple intersection attack unless the groups are very large
%\begin{enumerate}
%\item Current background knowledge models are insufficient to express independent releases practically.
%\end{enumerate}
%
%\item Randomized methods provably resist arbitrary auxiliary information
%\begin{enumerate}
%\item Results of many specifications analyses can be released with little noise BDMN, DMNS, NRS07, MT07
%\item For experimental examples look at EGS frequentset mining
%\end{enumerate}
%
%\end{enumerate}

\end{document}